\documentclass[aps,twocolumn,showpacs,superscriptaddress,prc]{revtex4}

\usepackage{epsfig}

\usepackage{color}

\usepackage{graphicx}
\begin{document}
\newcommand{\tb}{ {\bf {t}}}

\newcommand {\Co}{$^{57}$Co}

\newcommand {\Ar}{$^{39}$Ar}

\newcommand {\Na}{$^{22}$Na}

\newcommand {\fP}{$F_{prompt}$}

\newcommand {\mus}{$\mu$s}

\newcommand {\Leff}{$0.25$}

\newcommand {\LeffErr}{$0.25 \pm 0.01 + 0.01$(correlated)}

\newcommand {\SE}{$L_{\mathrm{eff}}$}

% You should use BibTeX and apsrev.bst for references
%\bibliographystyle{/usr/share/texmf/bibtex/bst/revtex4/apsrev}
%\usepackage{natbib}

%\preprint{}

\title{Measurement of scintillation efficiency for nuclear recoils in liquid argon}
% Optional argument for running titles on pages
%\title[]{}



\author{Dan Gastler}
\author{Ed Kearns}
\affiliation{Boston University, Boston, MA}
\author{Andrew Hime}
\author{Laura C. Stonehill}
\author{Stan Seibert}
\affiliation{Los Alamos National Laboratory, Los Alamos, NM}
\author{Josh Klein}
\affiliation{University of Pennsylvania, Philadelphia, PA}
\author{W. Hugh Lippincott}
\author{Daniel N. McKinsey}
\author{James A. Nikkel}
\affiliation{Yale University, New Haven, CT}


\date{\today}

\begin{abstract}

The scintillation light yield of liquid argon from nuclear recoils relative to
electronic recoils has been measured as a function of recoil energy from 10
keVr up to 250 keVr at zero electric field.  The scintillation efficiency, defined as the ratio of the nuclear recoil scintillation response
to the electronic recoil response, is \LeffErr~above 20 keVr. 

\end{abstract}

%\linenumbers

%\pacs{}
%\keywords{}

\maketitle


%%%%%%%%%%%%%%%%%%%%%%%%%%%%%%%%%%%%%%%%%%%%%%%%%%
\section{Introduction}
\label{sec:intro}

A number of existing and proposed experiments use liquefied noble gases as
detection media for Weakly Interacting Massive Particles
(WIMPs)~\cite{Aprile:2005a, Cline:2003, Brunetti:2005, Benetti:2007,
  Rubbia:2006}, a well
motivated dark matter candidate~\cite{Jungman:1996}. 
Liquefied noble gases have a high scintillation
yield, are relatively simple to purify of both radioactive contaminants and
light absorbers, and should be easily scalable to the large masses required
for very sensitive detectors. Although the best limit for the
spin-independent WIMP-nucleon cross section is currently set by the germanium-based
CDMS experiment~\cite{Ahmed:2009} at $3.8 \times 10^{-44}$ cm$^2$ for a 70-GeV
WIMP mass, the XENON-10 experiment has set
a comparable limit of $8.8 \times 10^{-44}$ cm$^{2}$ for a 100-GeV WIMP
mass~\cite{Angle:2007}, showing that liquefied noble gases are viable dark
matter targets. 

Events in a noble liquid dark-matter detector may arise from scattering off of 
the nucleus or atomic electrons; dark matter will only scatter off the nucleus 
to an appreciable extent.  The ratio of the scintillation light yield for 
nuclear recoil events relative to electronic recoil events is defined as the 
scintillation efficiency or \SE.

A WIMP dark matter search requires an energy threshold on the order of tens of keV, and it is necessary to measure the scintillation efficiency down to this energy threshold so as to quantify the WIMP detection sensitivity.
In order to make this measurement of \SE, a D-D neutron generator was used to produce neutrons that
scattered from a liquid argon detector into an organic liquid scintillator
detector used as a coincidence trigger.  The organic scintillator was placed
at a series of known angles, and the energies of the selected nuclear
recoils in the liquid argon were kinematically determined.  
The scintillation efficiency was determined from the ratio of
the measured electron-equivalent recoil energy at a given scattering angle to
the expected nuclear recoil energy (keVr) at that angle.  
Details of this measurement in a 4-kg liquid argon detector are presented in this paper, along with
scintillation efficiency results for nuclear recoil energies between 10 and 250
keVr at zero electric field. 

\section{Review of Physical Processes and Measurements}
Discrimination between nuclear recoil events that characterize a WIMP signal
and electronic recoil events that characterize the primary backgrounds is
essential in WIMP detectors, particularly for the case of liquid argon which
contains the radioactive isotope $^{39}$Ar. 
The noble liquid detectors use two methods to achieve this discrimination. 
Single-phase detectors use pulse-shape discrimination (PSD) based solely on
scintillation light to discriminate between event types, while dual-phase detectors can 
collect both scintillation light and ionization, employing a combination of PSD
and the relative size of the light and ionization channels to identify events. 
PSD is made possible because ionizing radiation in liquid noble
gases results in the formation of excited diatomic molecules (excimers) that
can exist in either singlet or triplet states, with very different lifetimes. 
In liquid argon these lifetimes are 7 ns and 1.5~$\mu$s, respectively \cite{Hitachi:1983, Lippincott:2008} 
and the scintillation light is produced in the decay of these states. 
As different types of excitation produce different ratios of triplet to singlet molecules, the relative
amplitudes of the fast and slow components can be used to determine what type of 
excitation occurred.
The effectiveness of this PSD is directly dependent on the number of detected photoelectrons in an event, and thus the 
light yield for both nuclear recoils and electronic recoils sets the energy
threshold for which electronic recoil backgrounds are negligible, in turn
determining the ultimate sensitivity of the detector to dark-matter-induced nuclear recoils. 

The excimers that provide the scintillation light are formed in two ways.  
An excited atom (exciton) can combine with another atom in the liquid to produce 
the excimer, or an ionized atom can combine with another atom to 
form a diatomic ion, which in turn recombines with an electron,
eventually resulting in the production of the excimer.  
The ratio of exciton production to ion pair production in liquid argon has 
been calculated to be 0.21 \cite{Takahashi:1975}, indicating that the majority 
of the scintillation light in liquid argon comes from excimers formed 
indirectly from argon ions, rather than directly from excited argon.  The 
average energy 
required to produce an electron-ion pair in liquid argon 
has been measured to be $23.6 \pm 0.3$ eV \cite{Miyajima:1974}, and 
the average energy needed to produce a single photon has been calculated 
to be $19.5 \pm 1.0$ eV \cite{Doke:2002}.  From this, the maximum 
possible scintillation yield in liquid argon is about 51 photons per keV of 
deposited energy, in the extreme case where the excimer formation and
scintillation processes are perfectly efficient.    

In actuality, the absolute light yield is reduced through a number of 
different mechanisms.  Energy may be lost by means other than exciton and 
ion pair formation, the excitons may undergo non-radiative collisions, and the
recombination of diatomic ions may be incomplete. The first mechanism is known
to be significant for nuclear 
recoils, for which a significant portion of the energy is lost to atomic 
motion as described by Lindhard \cite{Lindhard:1963}.  Thus, the 
scintillation light yield is expected to be reduced for nuclear recoil events 
compared to electronic recoils. 
Measurements of the scintillation efficiency for nuclear recoils relative to 
electronic recoils in liquid xenon \cite{Sorensen:2008ec,Aprile:2008rc,Aprile:2005, Chepel:2006,
  Manzur:2009} indicate 
that there is an additional reduction in the nuclear recoil scintillation
yield 
due to collisions between free
excitons that result in an ion and a ground-state atom, as described by 
Hitachi~\cite{Hitachi:1992}.  The rate of these biexcitonic collisions is 
dependent on the density of the excitations, thus the amount of quenching 
increases with increasing linear-energy-transfer (LET) and is most
significant at larger recoil energies. 
This mechanism of 
biexcitonic quenching is expected to apply to liquid argon as well as liquid
xenon, and a model for scintillation efficiency in argon, neon and xenon
taking LET into account has been proposed by Mei et al.~\cite{Mei:2008} 
A further reduction in scintillation yield can result when
some fraction of the ion-electron pairs do not 
recombine to produce an excimer and the electrons escape
instead~\cite{Doke:2002}. 

Relative scintillation efficiencies in liquid argon have been measured for a
number of different particle types.  For heavy fission fragments with kinetic
energy around 80 MeV, the scintillation efficiency relative to 1-MeV electrons
has been measured to be $0.21 \pm 0.04$~\cite{Hitachi:1982}.  
For alpha particles, the
scintillation efficiency has been measured to be 0.9 for 6-MeV alphas relative
to 1-MeV electrons~\cite{Hitachi:1982} and 0.4 for 5.3-MeV alphas relative to
1.2-MeV electrons~\cite{Regenfus:2007}.  
The observed triplet lifetime from the 5.3-MeV alphas was 800~ns, which
may indicate additional absorption due to impurities, but both values are 
plausible given the expectation that the scintillation efficiency for alphas 
should fall somewhere between that of fission fragments and unity.  
A previous measurement of scintillation efficiency for nuclear recoils in liquid
argon by the WARP collaboration gives $0.28 \pm 10\%$, measured at 65 keV 
average recoil energy~\cite{Brunetti:2005}.

%%%%%%%%%%%%%%%%%%%%%%%%%%%%%%%%%%%%%%%%%%%%%%%%%%%
\section{Experimental Details}
\label{sec:exp_det}

\begin{figure}[htbp]
  \centerline{
%    \hbox{\psfig{figure=Figs/cell.eps,width=8cm,%
    \hbox{\psfig{figure=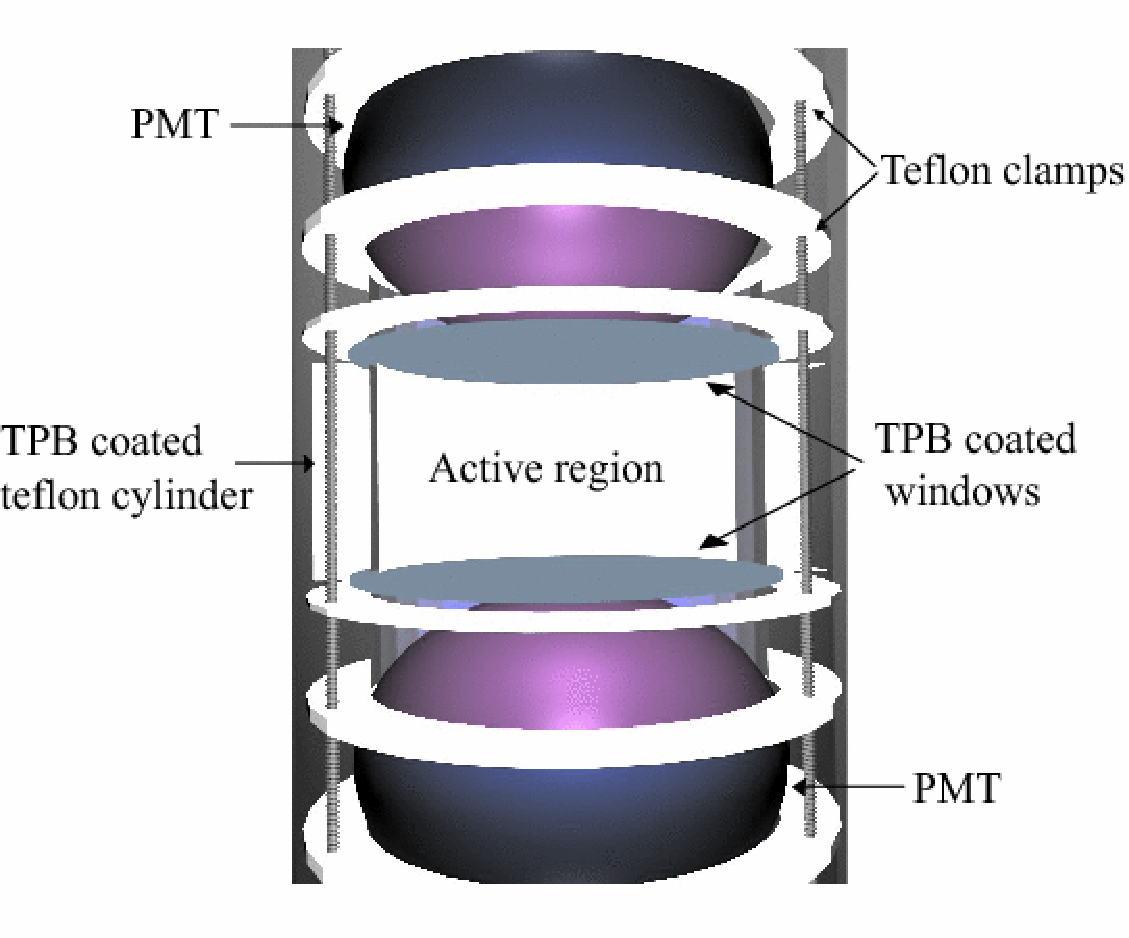,width=8cm,%
                clip=}}}
          \caption[MicroCLEAN scintillation cell]
                  {Schematic representation of the scintillation volume and PMT orientation.}
          \label{fig:cell}
\end{figure}

The liquid argon scintillation efficiency was measured using the MicroCLEAN detector at Yale University.
The active volume is 3.14 liters of liquid argon viewed by two 
photomultiplier tubes (PMTs). Figure~\ref{fig:cell} shows a schematic of the central region and
PMTs. The active region is defined by a teflon cylinder 200~mm in diameter and
100~mm in height, with two 3-mm-thick fused-silica windows closing the top and
bottom. Two 200-mm-diameter Hamamatsu R5912-02MOD photomultiplier tubes (PMTs)
are held in place by teflon rings above and below the central volume and view
the active region through the windows. Because liquid argon scintillates in
the vacuum ultraviolet at 128 nm~\cite{Cheshnovsky:1972}, all inner
surfaces of the teflon and fused silica are coated with a thin film of tetraphenyl
butadiene (TPB)~\cite{McKinsey:1997}.  The TPB shifts the wavelength of the
ultraviolet light to approximately 440~nm so that it may pass through the
windows and be detected by the PMTs. Both windows are coated
with $0.20 \pm 0.01$ mg/cm$^2$ of TPB and the teflon cylinder is coated
with $0.30 \pm 0.01$ mg/cm$^2$ of TPB. The teflon cylinder,
windows and PMTs are all immersed directly in liquid argon, contained within a
25-cm-diameter by 91-cm-tall stainless steel vessel.  

The stainless steel vessel is housed inside a vacuum dewar, and liquid argon
is introduced though a tube on the top of the vessel. The argon is liquefied
from purified gas in a copper vessel mounted to the end of a Cryomech PT805
pulse-tube refrigerator. All components that come into contact with the gas or
liquid are baked to at least 60$^{\circ}$C, and the ultra-high-purity argon
gas (99.999\%) is passed through a heated Omni Nupure III gas-purification
getter before entering the vessel.  Outgassing can cause impurities to build
up in the detector, decreasing the light yield by quenching the argon excimers
or absorbing the UV scintillation photons.  To avoid signal degradation, the
argon is continually circulated through the getter and reliquefied at a rate
greater than 2.0 slpm. No reduction in signal was observed during the
run. PMT signals were ditigized using an 8-bit 500~MSPS waveform digitizer 
with each of the PMTs capturing both low-gain and high-gain waveforms. 
More details about the experimental apparatus, data acquisition, and purity 
measurements are available in~\cite{Lippincott:2008}.  

\begin{figure}[htbp]
  \centerline{
%    \hbox{\psfig{figure=Figs/trace_ex.eps,width=8cm,%
    \hbox{\psfig{figure=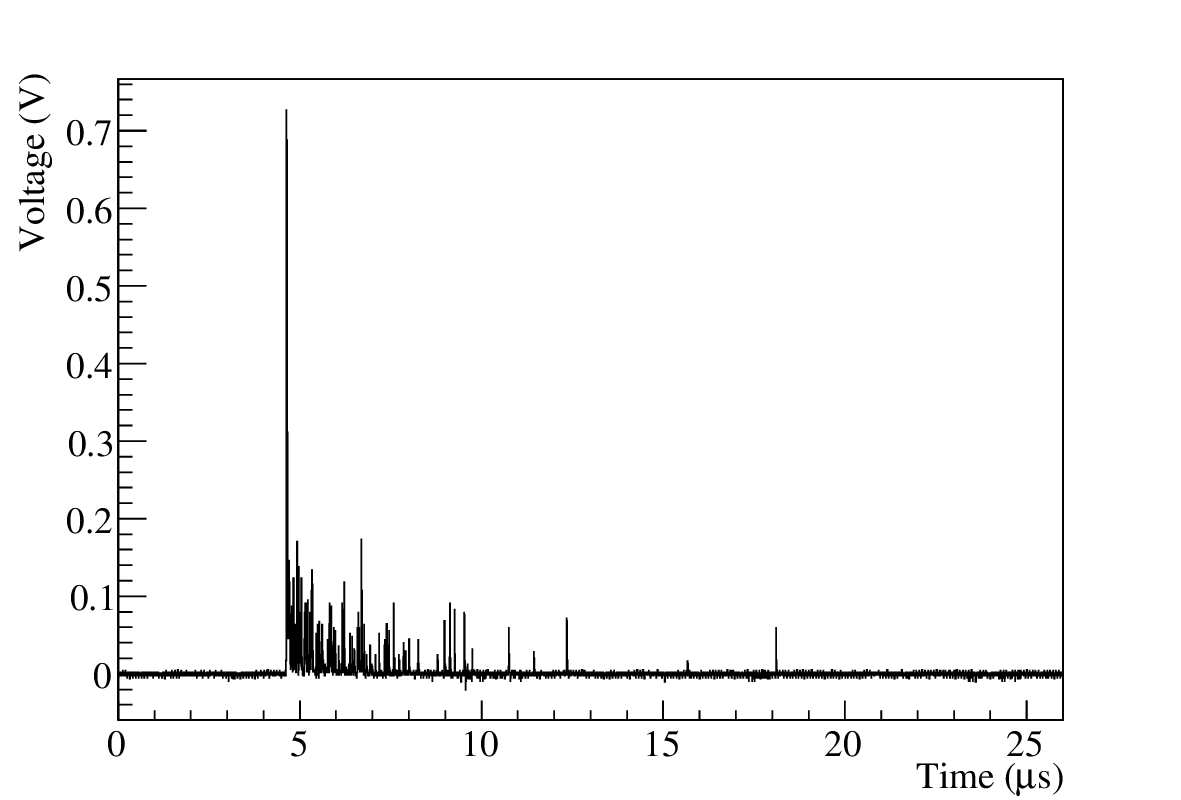,width=8cm,%
                clip=}}}
          \caption[Sample scintillation trace]
                  {Example of an electronic recoil-induced scintillation event in liquid argon.}
          \label{fig:trace_ex}
\end{figure}

\begin{figure}[htb]
  \centerline{
%    \hbox{\psfig{figure=Figs/calFig.eps,width=8cm,%
    \hbox{\psfig{figure=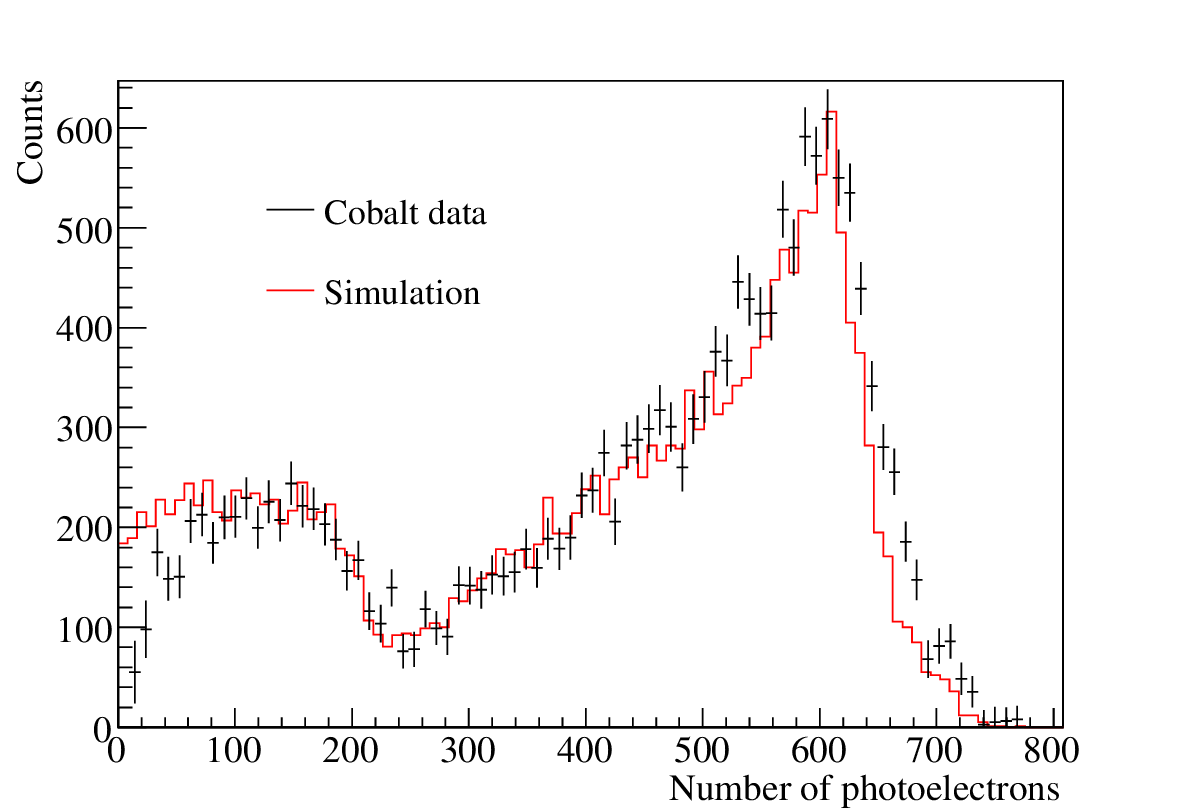,width=8cm,%
                clip=}}}
          \caption[Plot of $^{57}$Co data]
                  {(Color online) Plot of the $^{57}$Co spectrum, along with a simulation of the expected spectrum.}
          \label{fig:energyCalibration}
\end{figure}

A sample oscilloscope trace from an electronic recoil scintillation event in argon is shown in
Fig.~\ref{fig:trace_ex}. A 10-$\mu$Ci sealed $^{57}$Co source is used for
daily measurements of the scintillation light yield for electronic recoils,
with a sample spectrum shown in Fig.~\ref{fig:energyCalibration} along with
results from the simulation to be discussed in the next section. This source
produces 122-keV, 137-keV and 14.4-keV gamma-rays with branching ratios of
86\%, 11\% and 9\%, respectively.  Spectra were taken for each day of data taking,
with a Gaussian fit to the 122-keV peak providing a scintillation signal yield
calibration for that day in units of photoelectrons per keV of energy 
deposited by an electronic recoil, denoted as photoelectrons/keVee.  
Over the course of the four-month run, the signal yield remained 
stable to within 5$\%$ at $4.85 \pm 0.01$ photoelectrons/keVee. To check the
quality of the energy calibration, a 10-$\mu$Ci 
$^{22}$Na source that produces 511-keV gamma-rays is used as a second point of
reference, and the 511-keV line appears at a photoelectron yield that is
within 1\% of the value predicted from \Co\ source calibration.

\begin{figure}[htbp]
  \centerline{
%    \hbox{\psfig{figure=Figs/neutron_scatt_dia.eps,width=8cm,
    \hbox{\psfig{figure=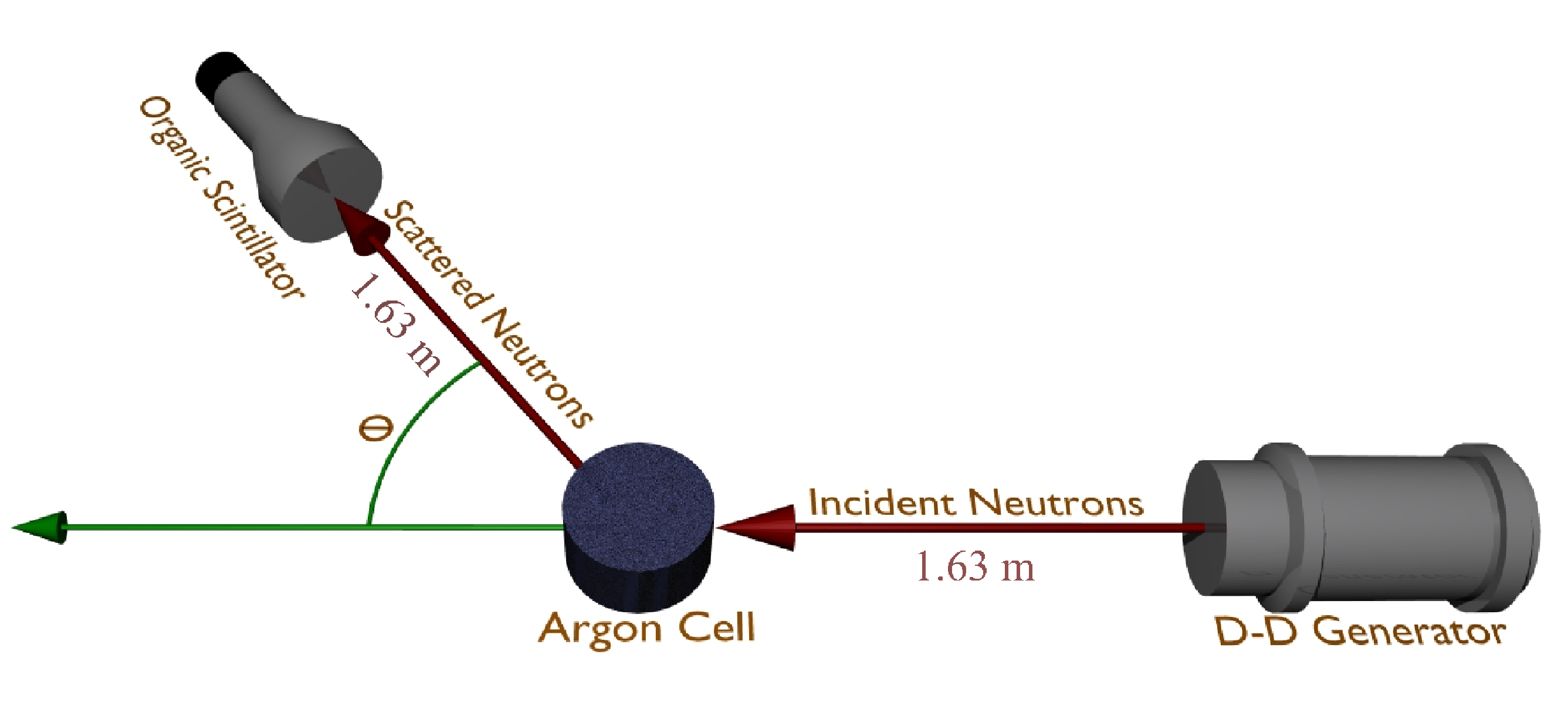,width=8cm,
                clip=}}}
          \caption[Neutron scattering schematic]
                  {(Color online) Top view of the neutron scattering setup.  Shown are
                  the neutron generator and the organic scintillator. The size
                  of the argon cell is not representative.  
                  }
          \label{fig:nuc_scatt}
\end{figure}

A portable Thermo Electron MP320 deuterium-deuterium (D-D) neutron generator
is used as a neutron source, with an organic scintillator detector as a
secondary coincidence trigger. The experimental setup can be seen
schematically in Fig.~\ref{fig:nuc_scatt}. In the forward direction, the D-D generator produces 2.8-MeV
neutrons. Some of these neutrons scatter in the liquid argon, and for a given
position of the organic scintillator, only neutrons that scatter at a specific
angle are selected by the coincidence trigger.  By changing the angle the
organic scintillator makes with the neutron generator-liquid argon detector 
axis, the scattering energy
of the recoil nucleus in the liquid argon can be varied according to the
following equation: 
\begin{equation}
\renewcommand{\arraystretch}{2}
\begin{array}{c}
E_{rec} = \frac{2 m_n E_{in}}{(m_n+m_{Ar})^2} \left[ m_n +  m_{Ar} - m_n\cos^2(\theta) \right. - \\ \left. \cos(\theta) \sqrt{{m_{Ar}}^2+{m_n}^2\cos^2(\theta)-{m_n}^2}  \right],
\end{array}
\label{eq:theta}
\end{equation}
\noindent where $E_{in}$ is the incident neutron energy (2.8~MeV), $A$ is the
atomic mass number, and $\theta$ is the scattering angle of the
outgoing neutron. Data were taken at
19 different angles corresponding to recoil energies between 10 keV and 250
keV. 
The setup also included 12 inches of poly between the neutron generator and the organic scintillator to reduce the accidental coincidence rate, although this is not shown in Figs.~\ref{fig:nuc_scatt} or~\ref{fig:MC_geo}. 
	
%%%%%%%%%%%%%%%%%%%%%%%%%%%%%%%%%%%%%%%%%%%%%%%%%%%
\section{Monte Carlo Simulations}
\label{sec:monte}

In order to understand the data, we developed a Monte Carlo simulation of the argon 
detector, cryostat, organic scintillator and surrounding lab space.  The software 
framework used was RAT, which combines 
Geant4\cite{GEANT}, CLHEP\cite{CLHEP}, and ROOT\cite{ROOT} into a single simulation and analysis package.  
A detailed optical model of the inner detector and PMTs is included in 
the Monte Carlo which allows us to estimate smearing of the detected signal.  
While this model gives results that are in fairly good agreement with our gamma 
calibrations, we add an addition smearing term for the neutron scattering analysis 
to take into account the lower photon yield for a given energy.

\begin{figure}[ht]
\centering
\includegraphics[width=8cm]{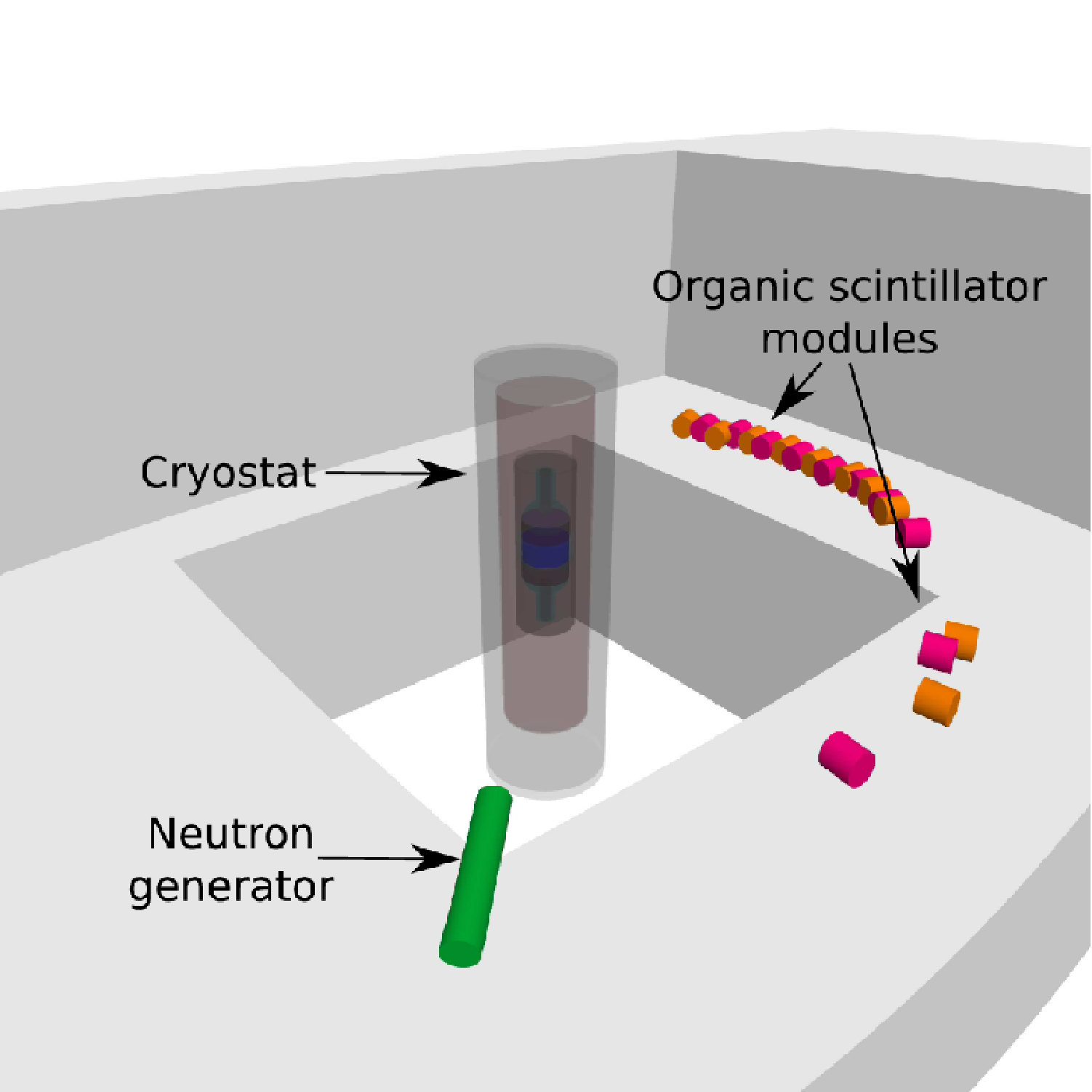}
\caption[Monte Carlo geometry]
  {(Color online) Monte Carlo detector geometry.}
\label{fig:MC_geo}
\end{figure}

An image of the detector geometry can be seen in Fig.~\ref{fig:MC_geo}.  
The argon detector is in the central vertical cylinder, where the various
layers of steel are set to be semi-transparent so that the inner workings are
visible.  While there was only one organic scintillator detector at
a given time in the real experiment, the simulation used many such detectors, 
so as to allow Monte Carlo data for half of the organic scintillator positions 
to be collected simultaneously and reduce computation time. In the picture, the cylinders 
off to the right of the argon detector represent the organic scintillator in 
its various positions.  As the adjacent positions of the organic scintillator 
overlap, independent simulations were performed for each set of colored 
organic scintillator positions. 
The 12 inches of poly between the generator and the organic scintillator locations was not included in the simulation.

In addition to the neutron scattering simulation, the detector response to an
external $^{57}$Co gamma source was also modeled.  The origin of the
122~keV and 137~keV gamma rays was set just outside of the outer vacuum
can as in the real detector.  The results of this simulation can be seen in
Fig.~\ref{fig:energyCalibration} showing very good agreement with the
experiment.

%%%%%%%%%%%%%%%%%%%%%%%%%%%%%%%%%%%%%%%%
\section{Analysis}
\label{sec:analysis}

The analysis begins by combining the high-gain and low-gain waveforms for each PMT in a given event. 
To do this the first two microseconds of each PMT's high-gain and low-gain waveforms are separately averaged to calculate baselines.
These baselines are then subtracted to give zero-offset waveforms. 
Each high gain waveform is then scanned to determine if the waveform digitizer is saturated. 
In the case of saturation, the high-gain and low-gain waveforms are aligned and the low-gain waveform's samples are inserted where the high-gain waveform was saturated. 

At this point each PMT waveform is scanned to determine the trigger time of the event. 
The trigger time for each PMT is defined as the time where the waveform reaches 20\% of its maximum height.
The average of the two trigger times is taken as the start time for the event and 
a timing cut is applied to the two PMT waveforms to remove events where the difference in timing is greater than 20 ns. 
The waveforms from both PMTs are then integrated in two timing intervals, the first from 20 ns before the trigger time to 100 ns after and the second from 100 ns to 5~$\mu$s. 
The region between 5~$\mu$s and 14~$\mu$s is scanned for single photoelectron pulses and used to determine the single photoelectron spectrum.
Any region where the waveform's voltage value exceeded approximately one third of a single photoelectron's peak voltage is integrated from 10 ns preceding the crossover sample to 50 ns following it.  
After this procedure has been performed on every event in a run, the run's single photoelectron value is fit and used to convert the integrated waveform charges into photoelectrons. 

A PMT asymmetry cut is used to remove events that are near the windows of the detector. 
The asymmetry is defined as the difference in the signals observed by the
two PMTs divided by their sum, and events with an asymmetry of more than 60\% are removed.
Events with approximately 2000 photoelectrons or more can cause the saturation of one or both of the detector PMTs, and since this will cause events to have poor energy reconstruction, a cut is applied to remove events in which either PMT's output becomes greater than 2 V. 
This cut removed nuclear recoil events with energies above 110~keVee for runs with recoil angle below $125^{\circ}$ degrees and nuclear recoil events with energies above 180~keVee for $125^{\circ}$ and $142^{\circ}$ degree runs.
All cuts applied to the data up to this point are considered to be data quality cuts and were applied to both {\Co} and neutron runs.

To distinguish neutron scatters from other backgrounds in the neutron data
sets, two additional cuts involving the organic scintillator are applied. 
The first is a time-of-flight (TOF) cut  which removes events in which the organic
scintillator is triggered before the detector, as well as events in which the neutron
arrives late due to multiple scatters. 
In addition, this cut also helps remove background gammas from our neutron data sets. 
The position and width of this cut is set by the location of the single 
scattering neutrons in the Monte Carlo TOF spectrum. 

\begin{figure}[htbp]
  \begin{center}
    \includegraphics*[width=8cm]{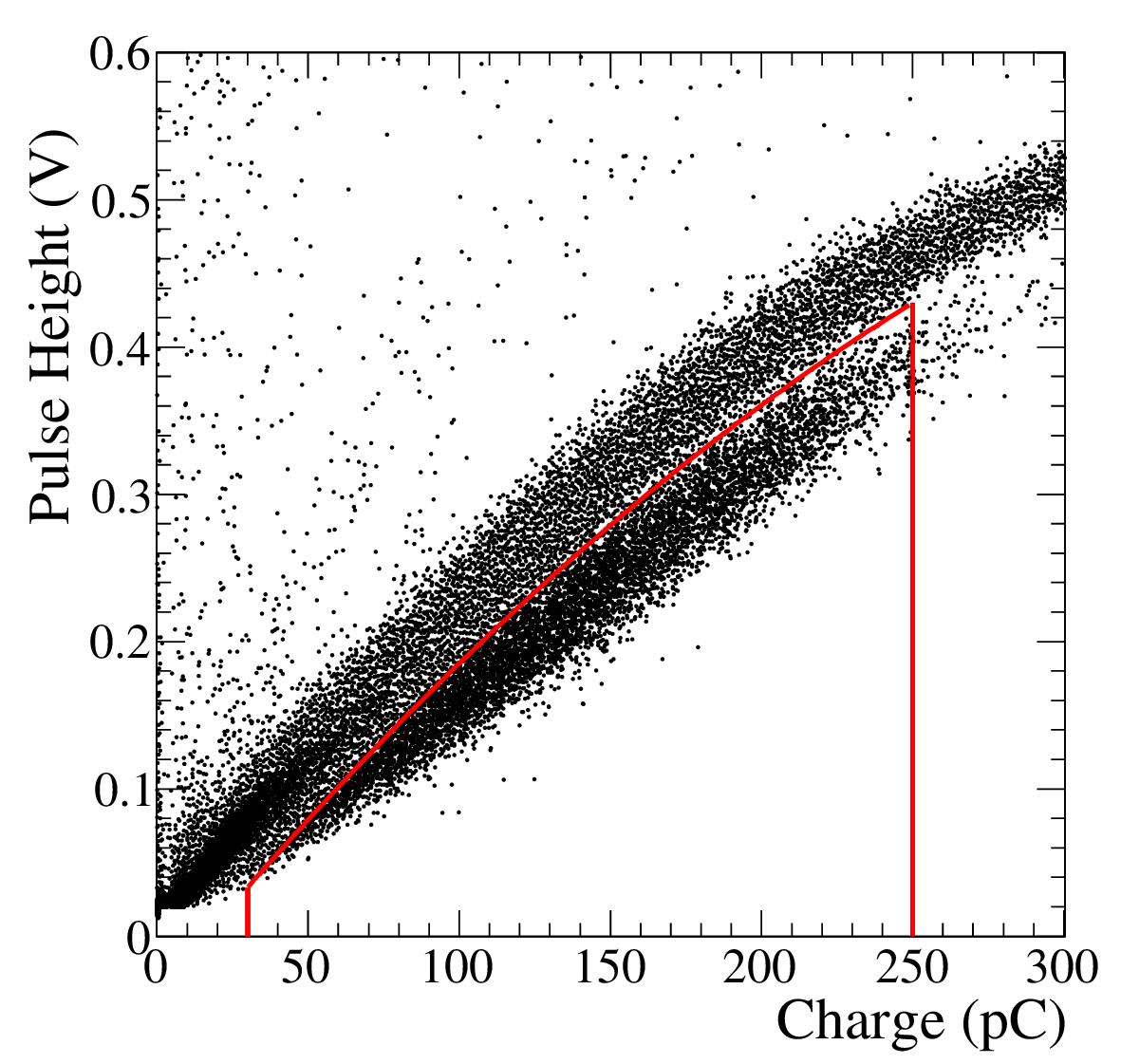}
    \caption{(Color online) An example of the pulse shape cut applied to the organic
      scintillator. A scatter plot of the pulse area versus peak pulse height shows a distinct region of neutron events bounded by the red quadratic curve and lines. } 
    \label{fig:OrgScinPSD}
  \end{center}
\end{figure}

The second cut associated with the organic scintillator uses the pulse shape
in the organic scintillator to distinguish between neutrons and background
events. 
We use a scatter plot of the pulse area within 100 ns of the organic 
scintillator trigger versus the organic scintillator waveform's maximum voltage,
shown in Fig~\ref{fig:OrgScinPSD}, to determine a quadratic curve that
separates two types of events.
This curve divides the scatter plot into two distinct
regions: an electronic recoil band that appears for all types of runs and a
nuclear recoil band that is only present in neutron runs.  
The separation of these regions did not appear to change with recoil energy and variation of the regions had a negligable effect on the results. 

We apply one final cut to remove electronic recoils from the sample,
exploiting the PSD to discriminate between event types in liquid argon. 
Based on our previous work~\cite{Lippincott:2008}, we define a discrimination parameter, \fP, as the
fraction of light arriving in a prompt time window. 
For comparison, the mean \fP~for recoils between $5$~and $32$~keVee range
from $0.39$~to $0.28$~for electronic recoils and $0.56$~and $0.7$~for nuclear recoils~\cite{Lippincott:2008}.
We apply a relatively loose cut, removing events with \fP~$ < 0.35$ from the final sample.
We checked the analysis with a tighter cut of \fP~$ < 0.50$ and incorporated the differences in the systematic uncertainties.

To determine the scintillation efficiency of the liquid argon, the measured
energy spectrum for each scattering angle is compared to the energy
spectrum produced by the Monte Carlo simulation analyzed with the same asymmetry and TOF
cuts. We also use the Monte Carlo simulation to calibrate the TOF cut. For each
scattering angle in the Monte Carlo simulation, a TOF spectrum is produced using
events that only scattered once in the detector before reaching the organic 
scintillator.  This allows us to find the range of the TOF for single scattering
neutrons for each position.  
After applying the same TOF and PMT asymmetry cut to the simulation as used for
the data, a Monte Carlo energy spectrum is then generated for each recoil angle.

There are two convolutions applied to the Monte Carlo recoil spectra before
fitting them to the data.  First, to account for the variation in the single
photoelectron charge, the Monte Carlo photon counts are smeared using the measured single
photoelectron charge distribution from the photoelectron calibration data. Second, since the
simulations were performed assuming a 100\% scintillation efficiency, an
additional smearing of $3.25 \times \sqrt{(1 - L_{\mathrm{eff}})N_{pe}}$ is applied
to the Monte Carlo to account for the difference in counting statistics between a
scintillation efficiency of 100\% and that obtained from the data.   
The proportionality constant of $3.25$ empirically account for the observed 
broadness of the clearly resolved peaks at 191, 211 and 239 keVr scattering 
angles. 
It is well known that noble liquid detectors do not reach the ideal energy 
resolution predicted by Poisson photoelectron statistics, largely due to 
ionization-scintillation anti-correlation\cite{Conti:2003, Aprile:2005a}.

The MINUIT fitting package is used to perform a $\chi^{2}$ fit of the 
Monte Carlo to the data with the normalization and the scintillation 
efficiency as free variables. 
Each Monte Carlo spectrum is binned using the same binning as
the corresponding recoil data and used to generate a spline for fitting.   
First, the entire range of the data is used in the fit. Then, a Gaussian is
fit to all events in the Monte Carlo identified as singly-scattered
neutrons. This fit is used to define a new fit range consisting of plus and
minus three sigma around the centroid of the Monte Carlo single scattered 
neutron distribution. The final fits are performed over a restricted range 
around the single scattering Monte Carlo neutron distribution where we 
expect to observe our signal. 
This fitting procedure was checked using the Monte Carlo sample with a set \SE~ value of $0.25$ and was able to recover this set \SE~ at each recoil angle.

\begin{figure*}[htbp]
  \begin{center}
    \includegraphics*[width=4.4cm]{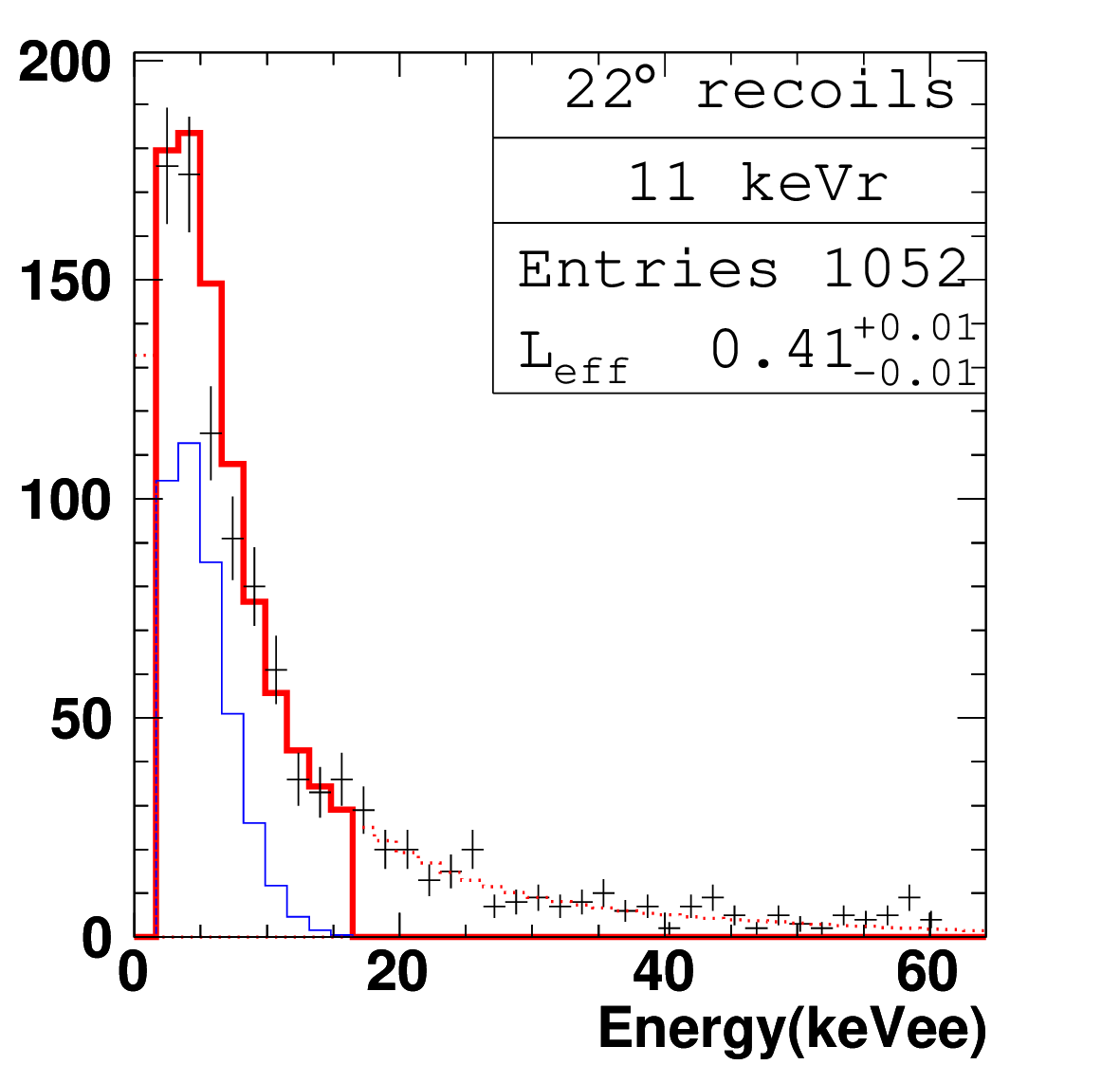}
    \includegraphics*[width=4.4cm]{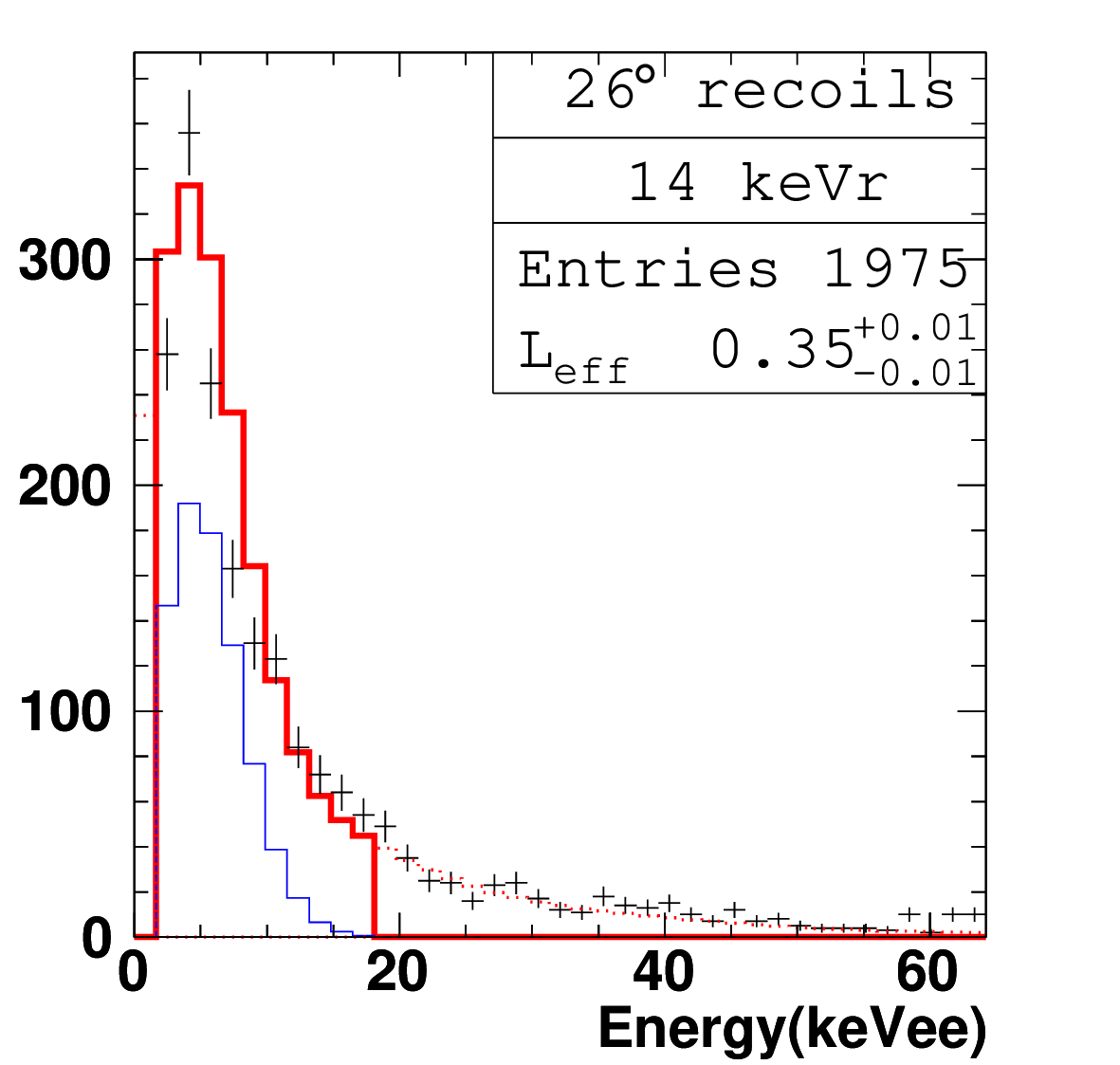}
    \includegraphics*[width=4.4cm]{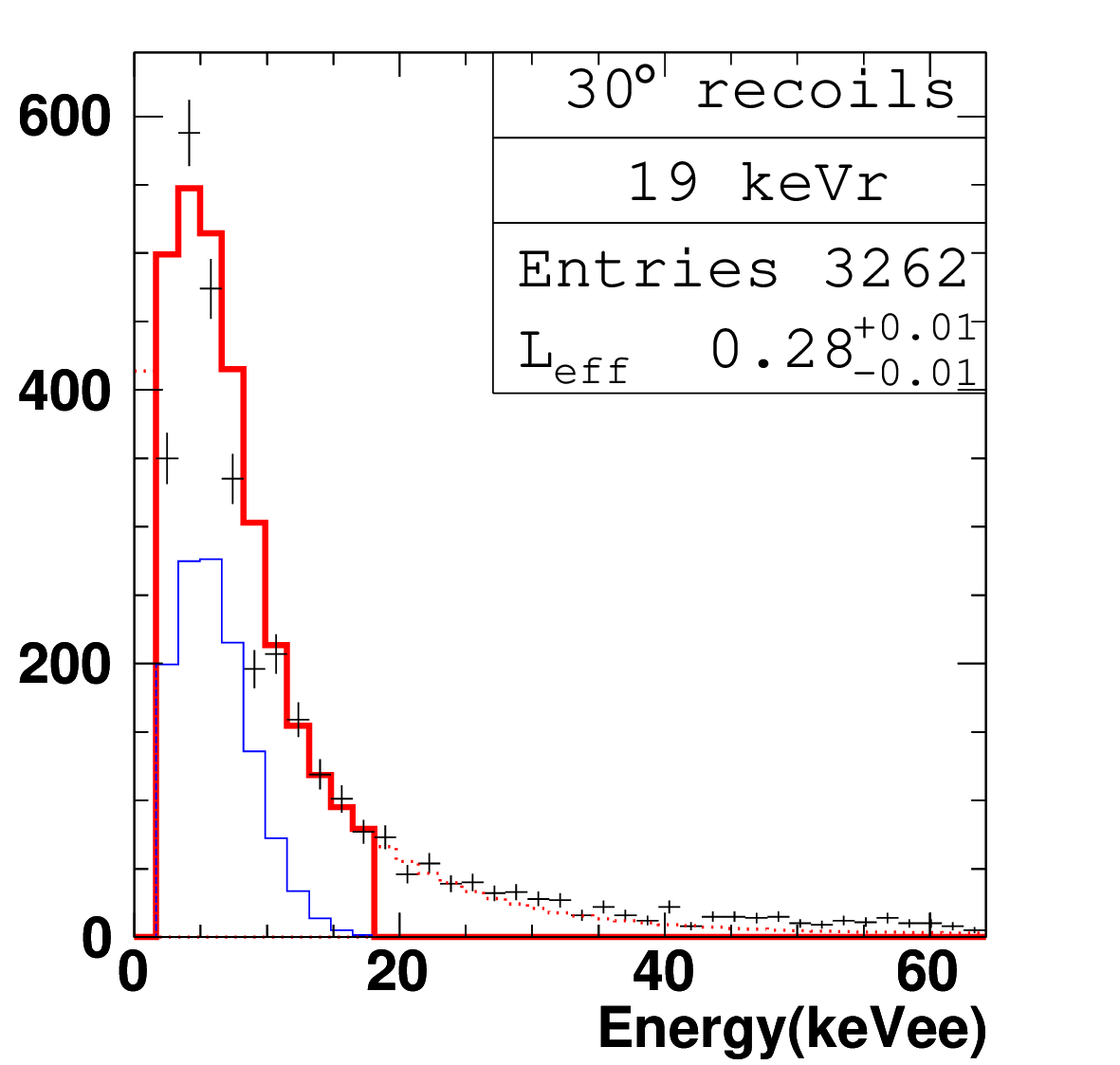}
    \includegraphics*[width=4.4cm]{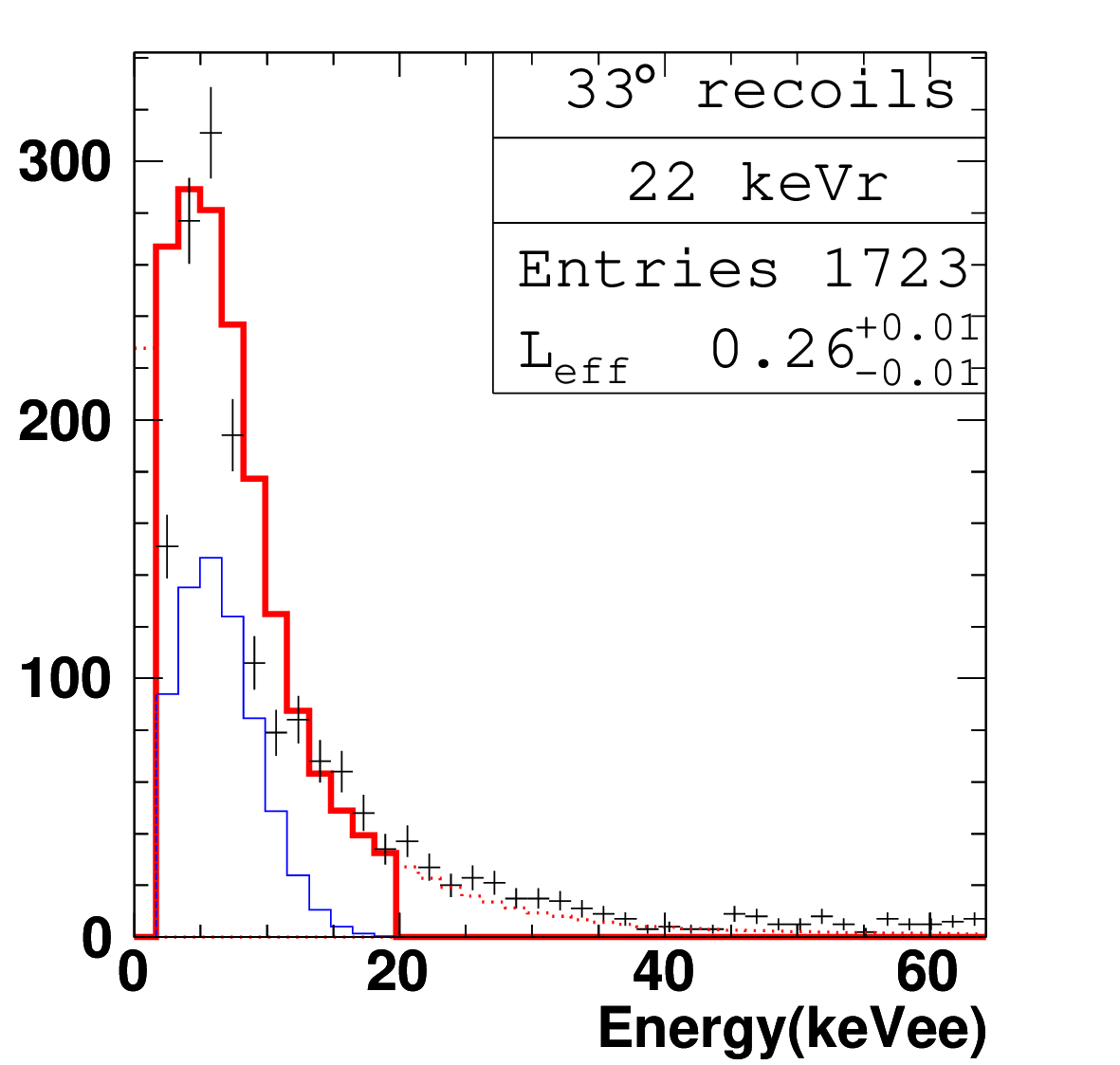}
    \includegraphics*[width=4.4cm]{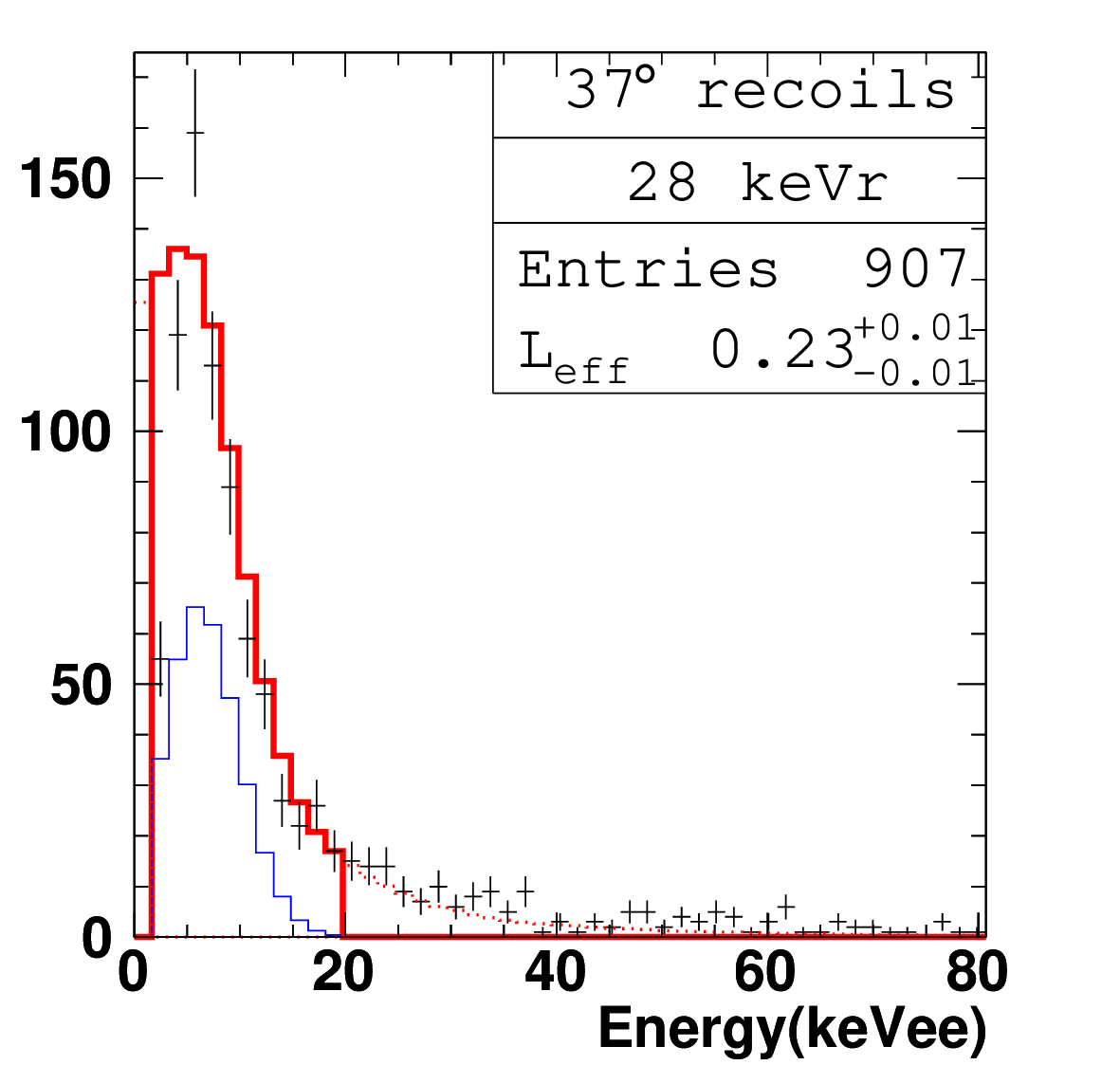}
    \includegraphics*[width=4.4cm]{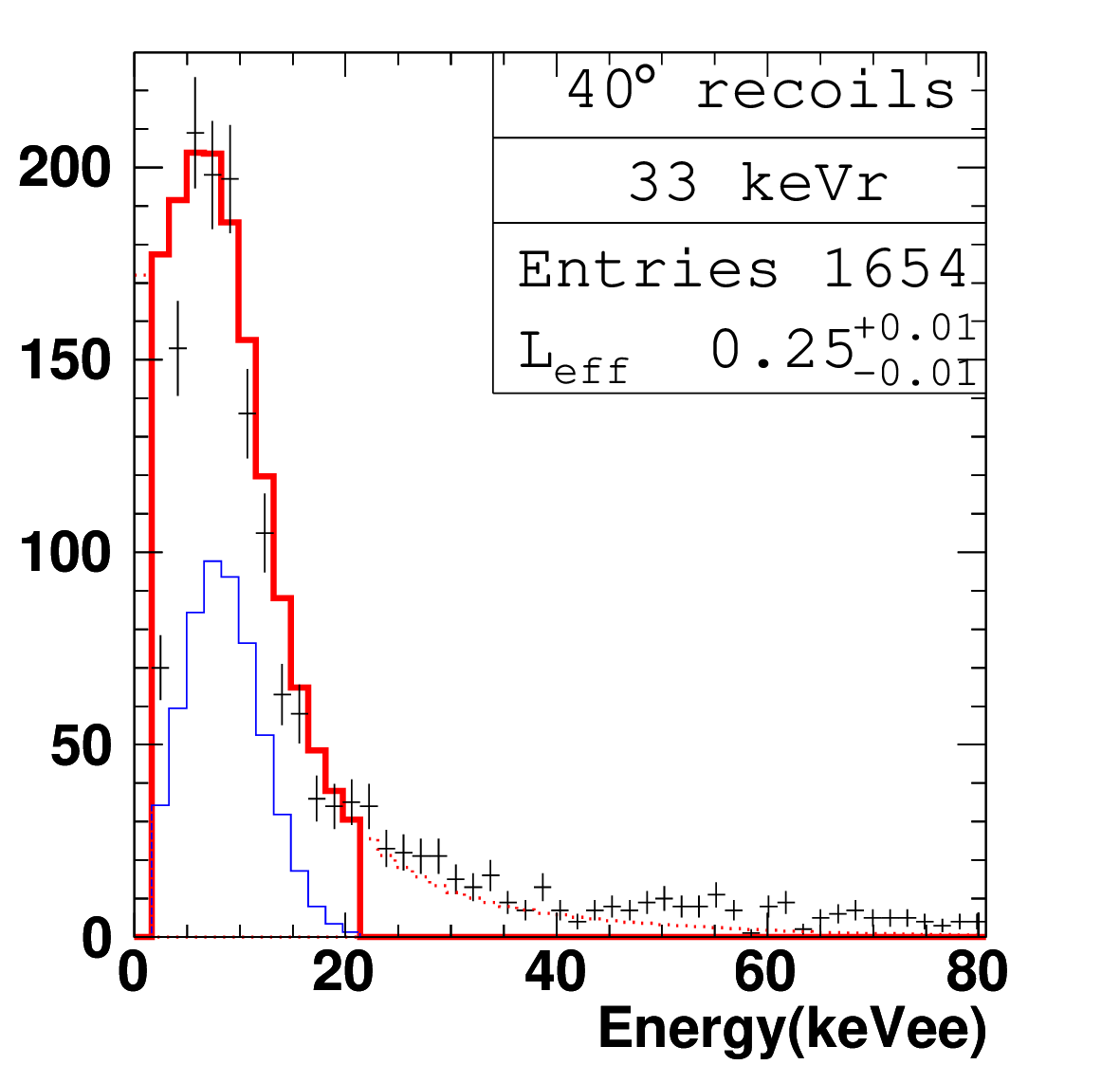}
    \includegraphics*[width=4.4cm]{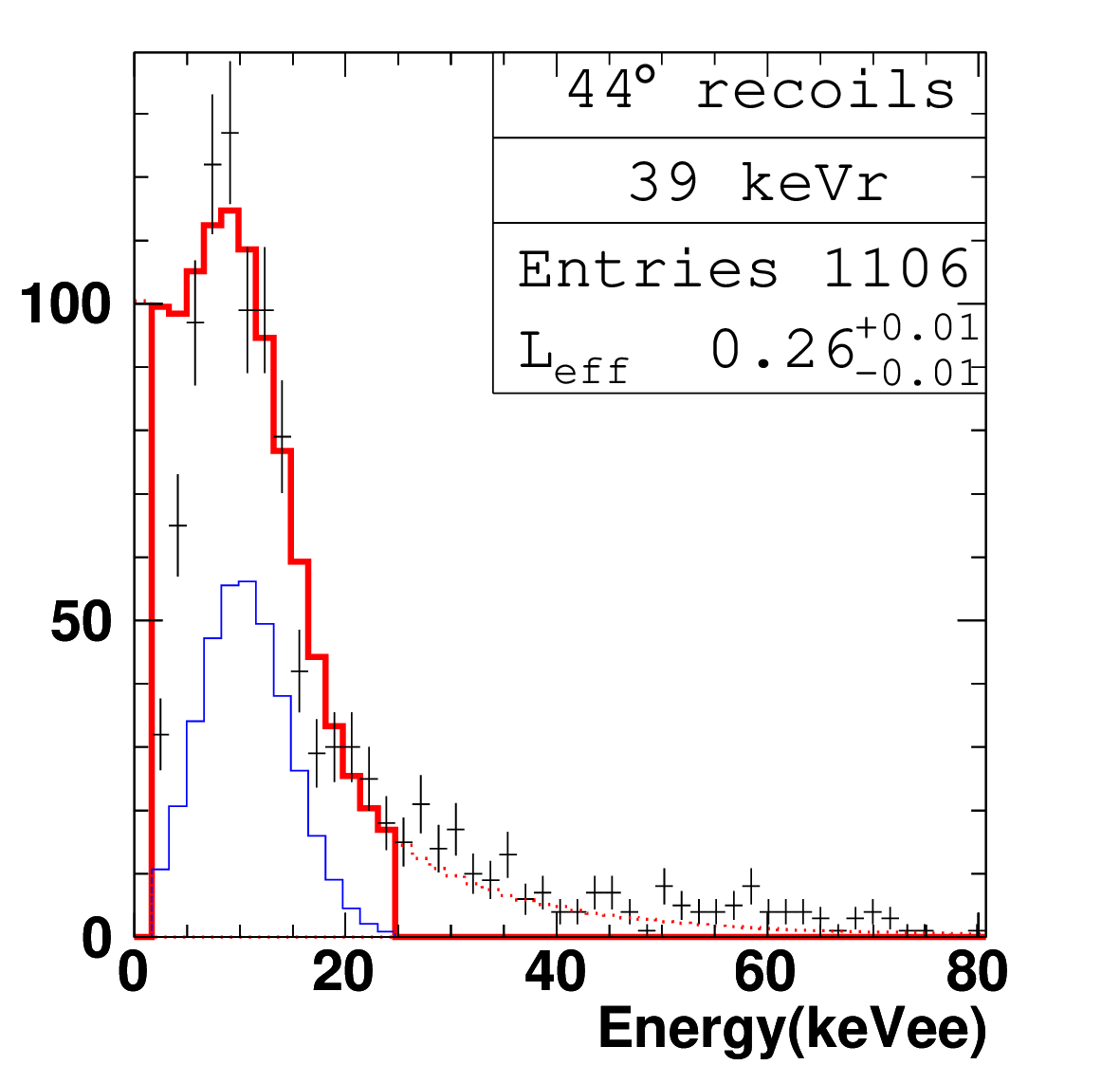}
    \includegraphics*[width=4.4cm]{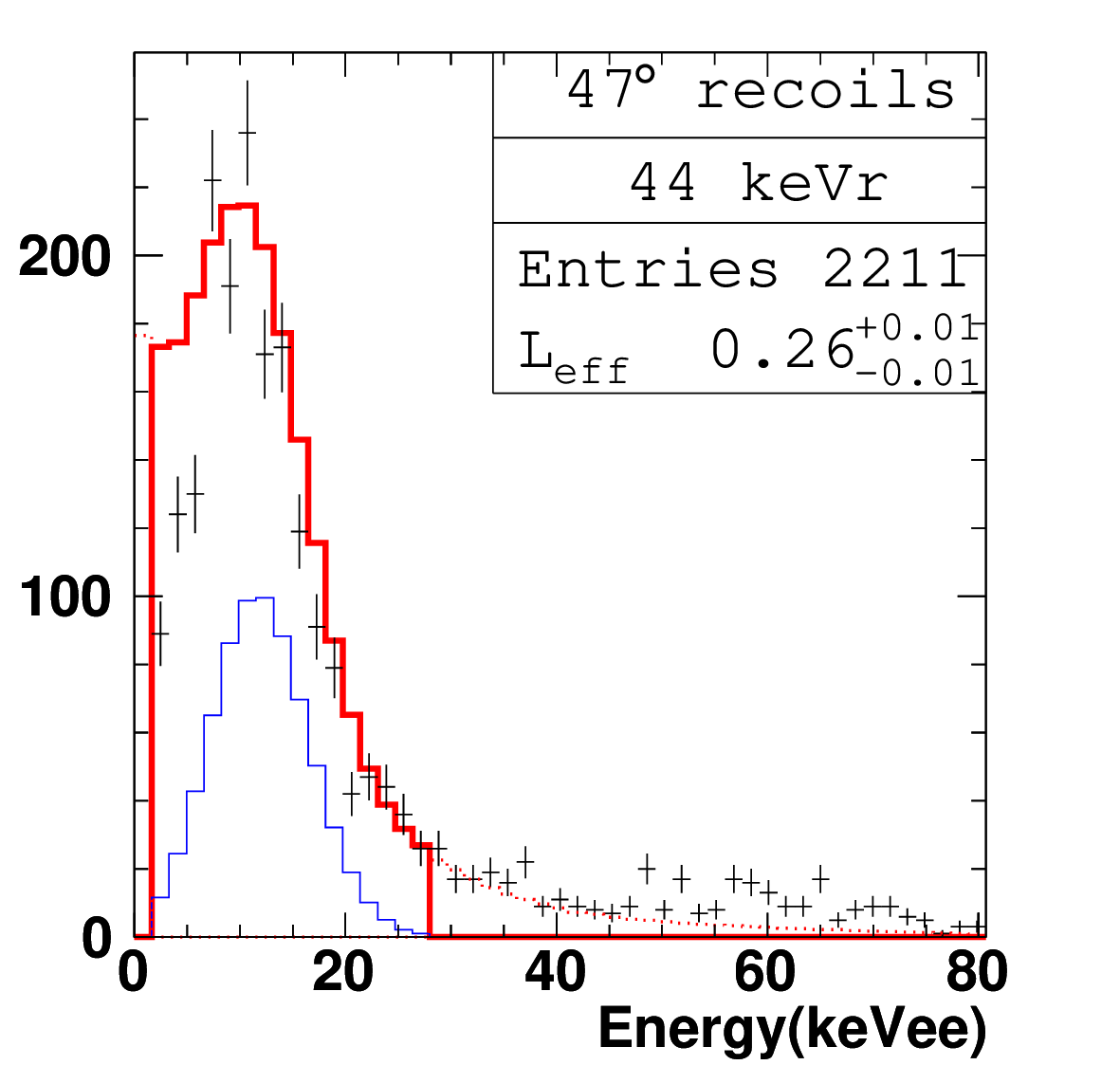}
    \includegraphics*[width=4.4cm]{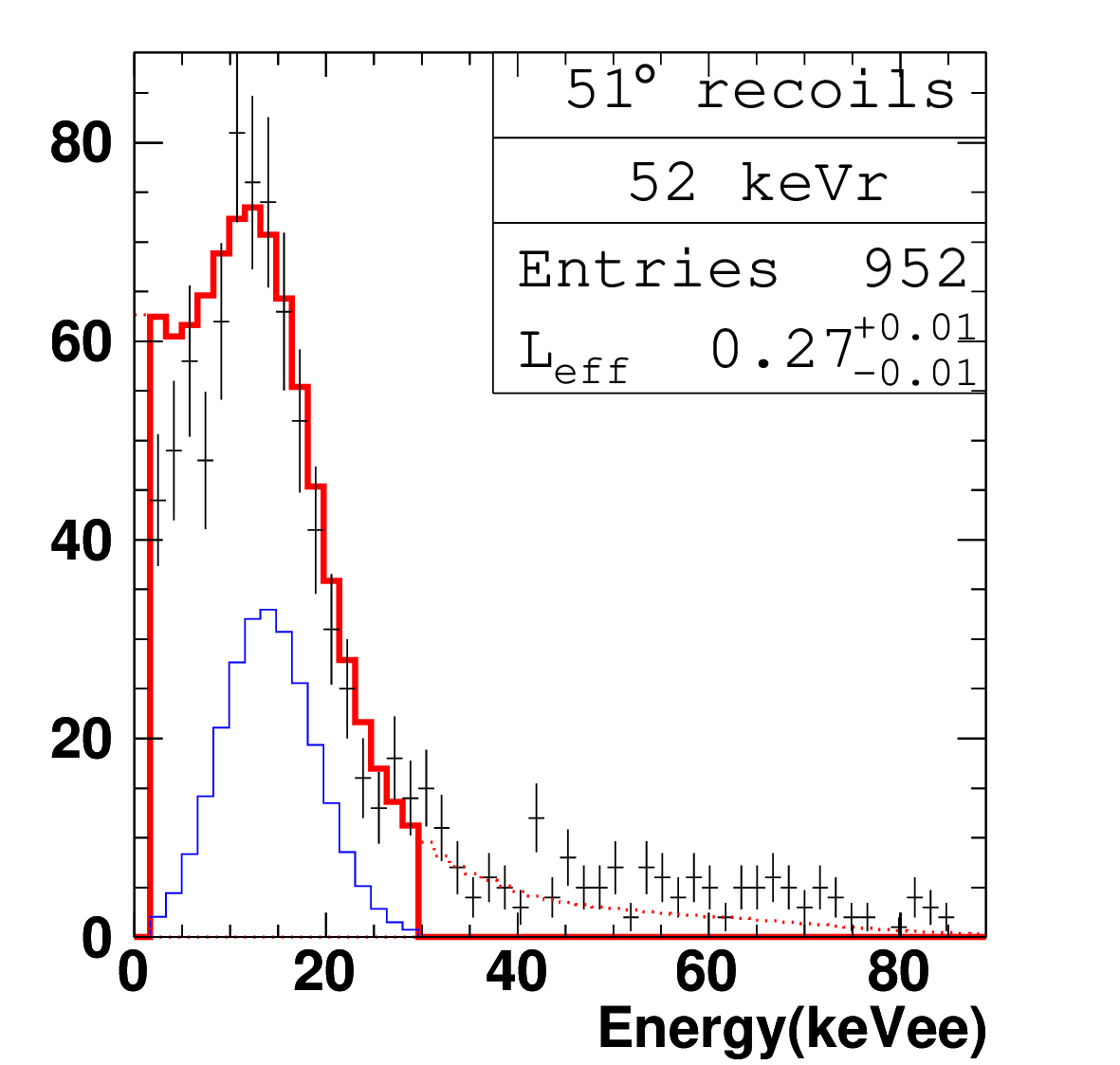}
    \includegraphics*[width=4.4cm]{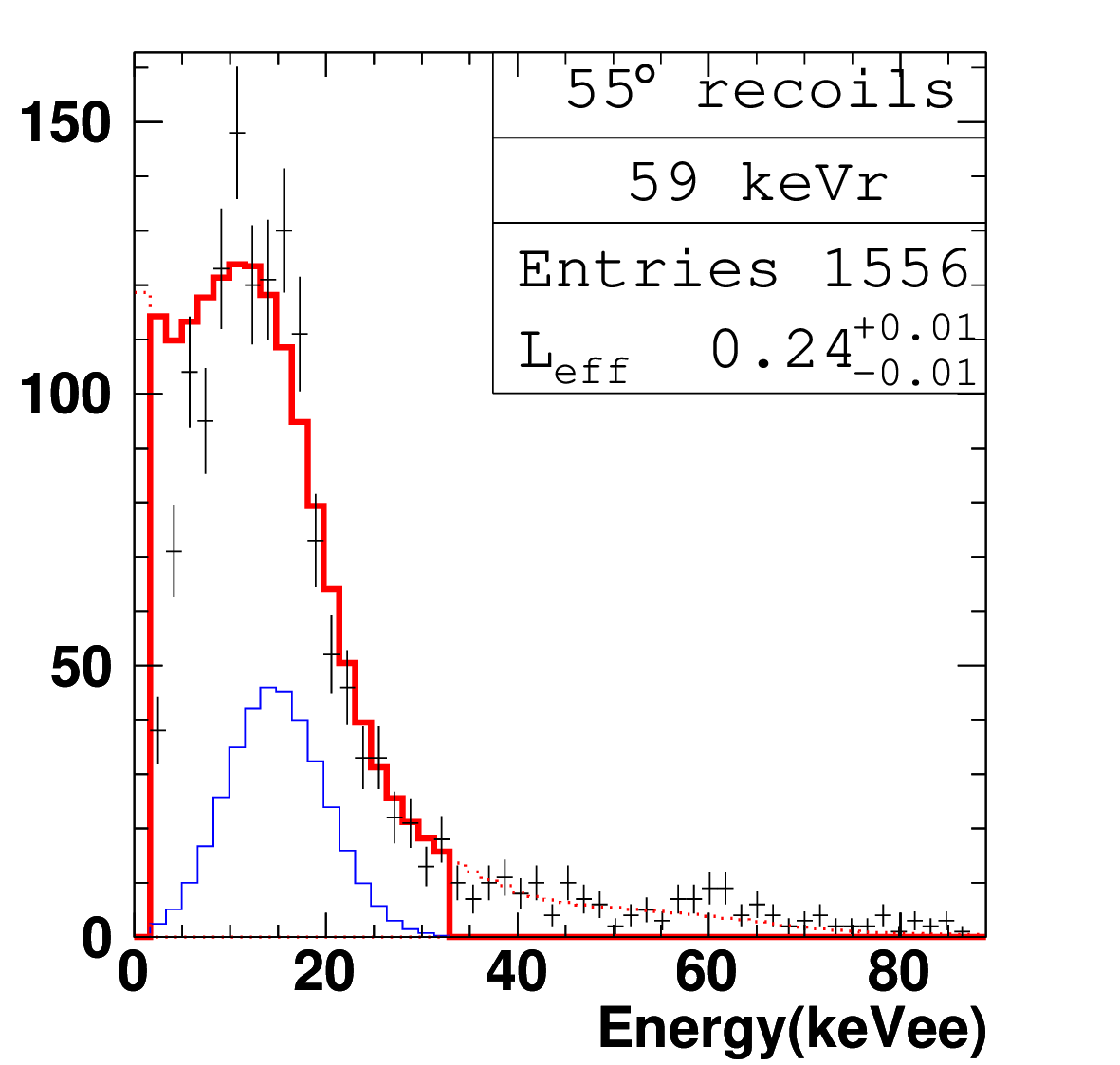}
    \includegraphics*[width=4.4cm]{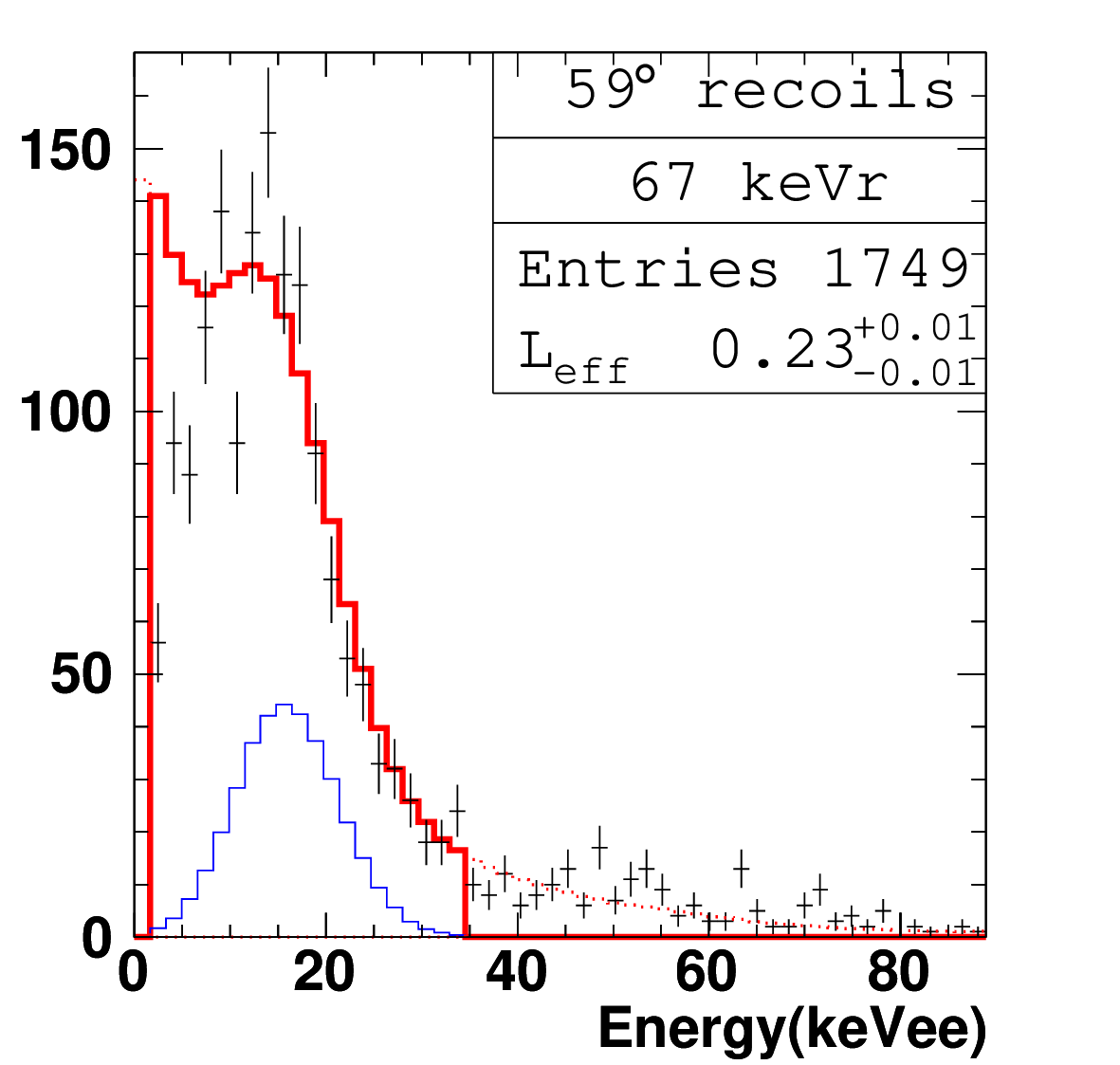}
    \includegraphics*[width=4.4cm]{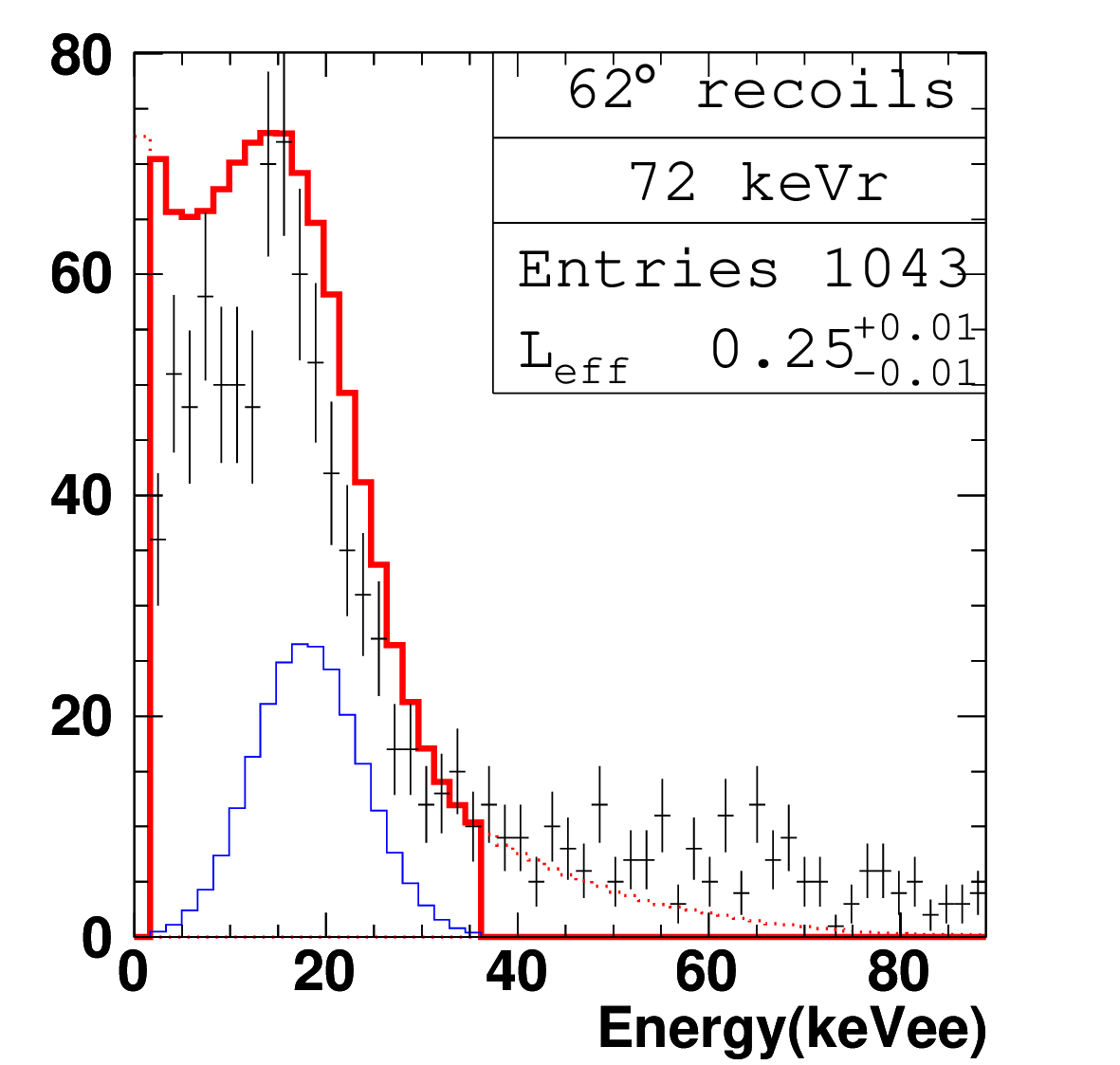}
    \includegraphics*[width=4.4cm]{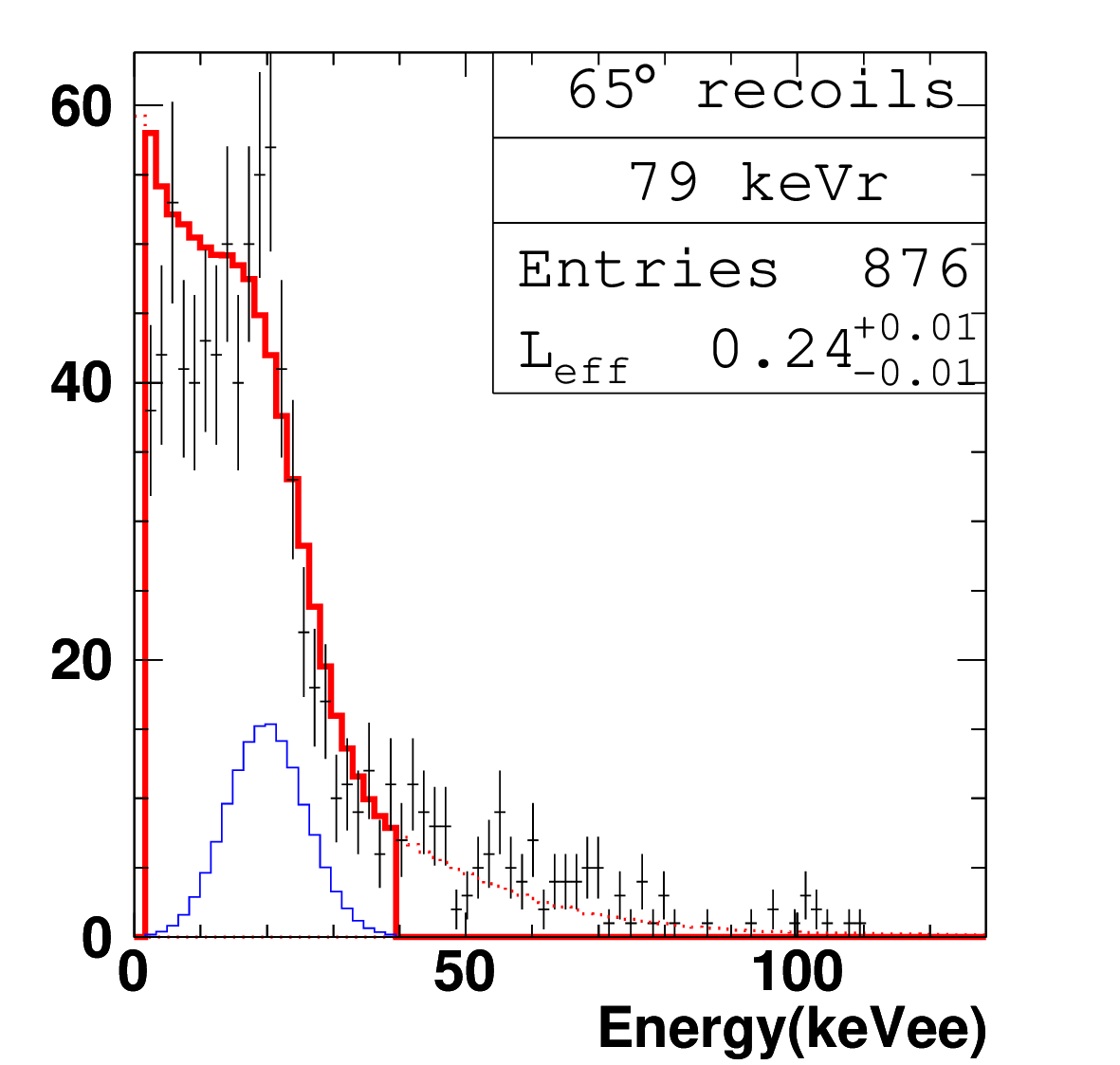}
    \includegraphics*[width=4.4cm]{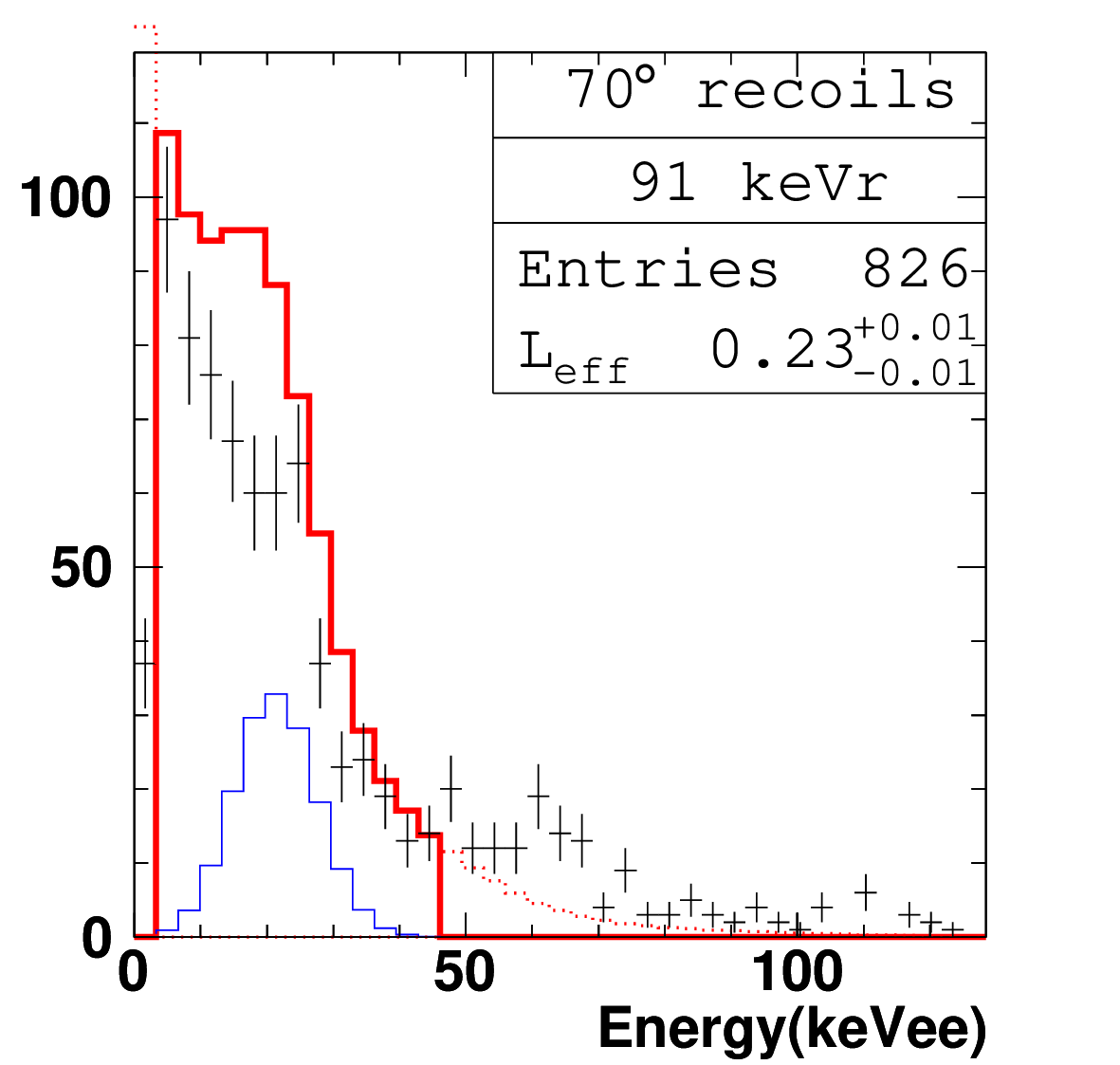}
    \includegraphics*[width=4.4cm]{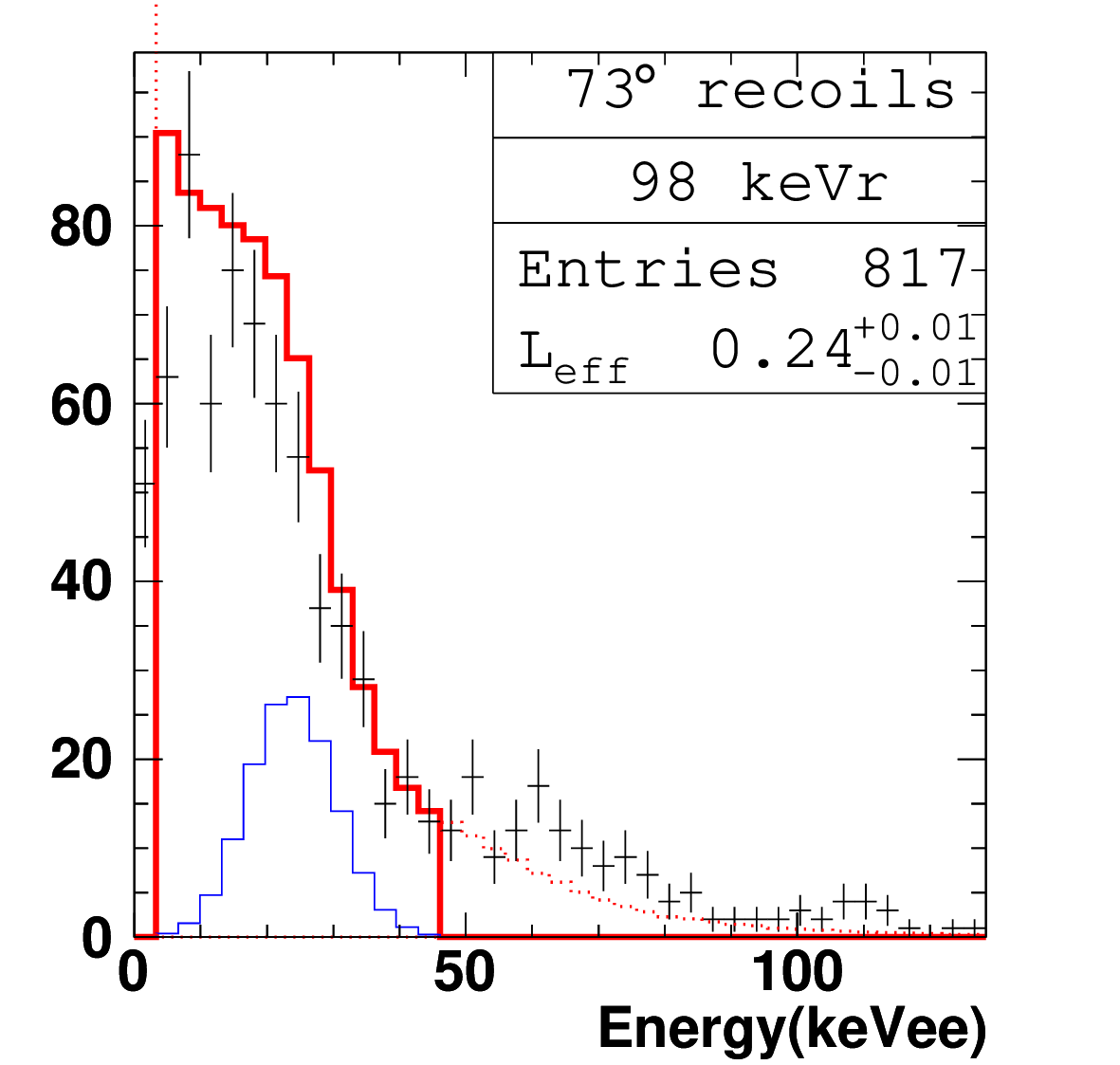}
    \includegraphics*[width=4.4cm]{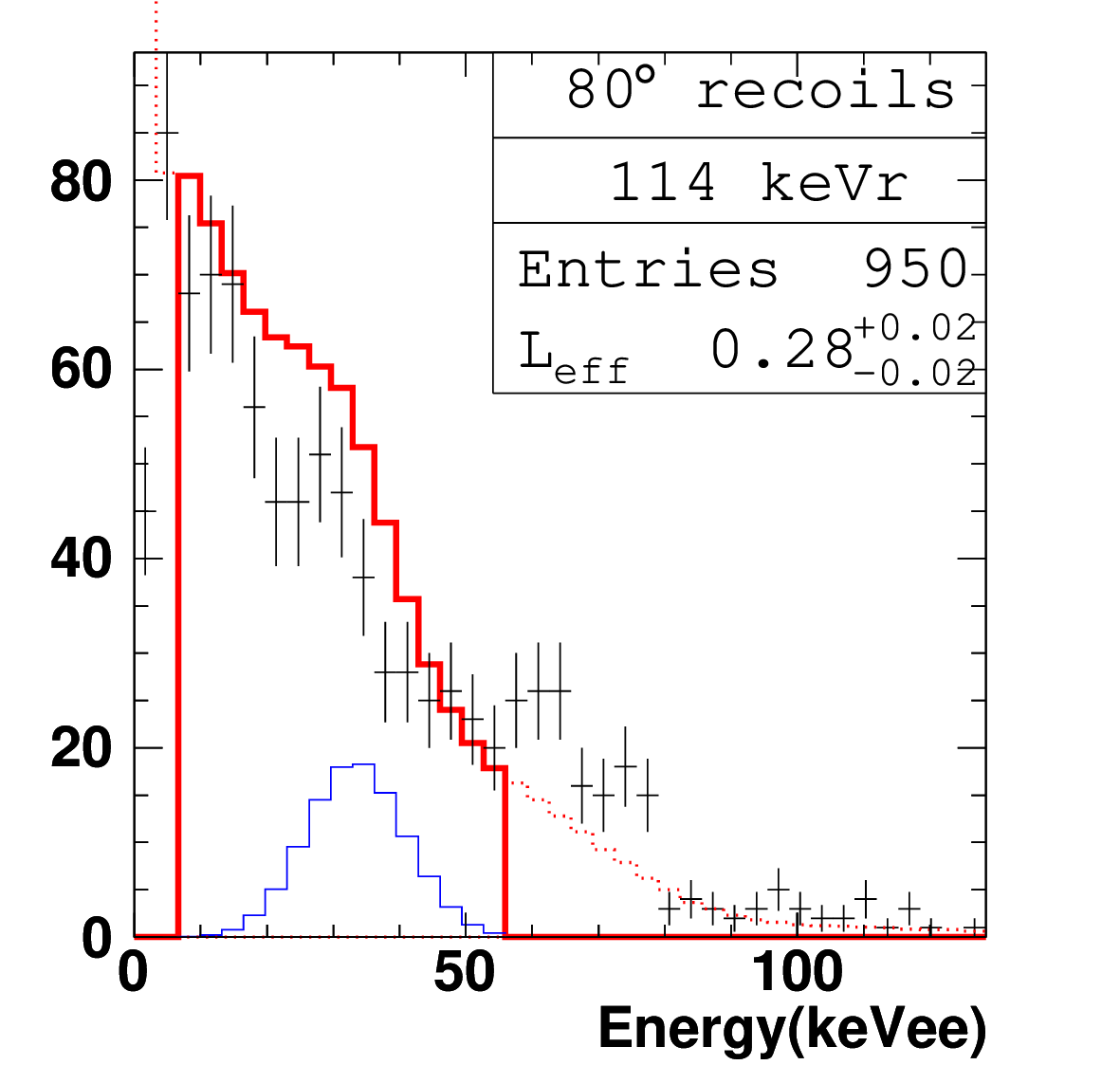}
    \includegraphics*[width=4.4cm]{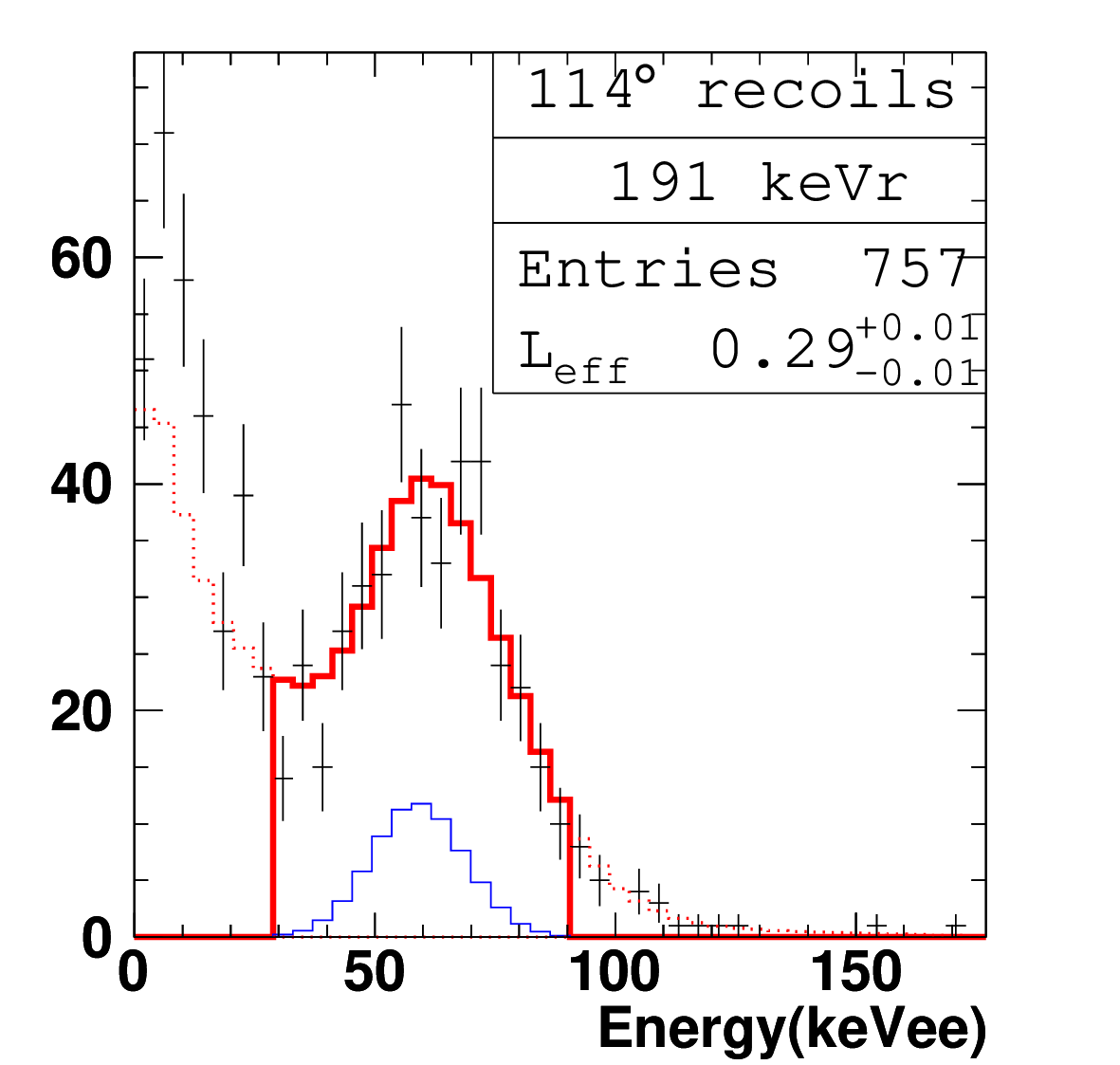}
    \includegraphics*[width=4.4cm]{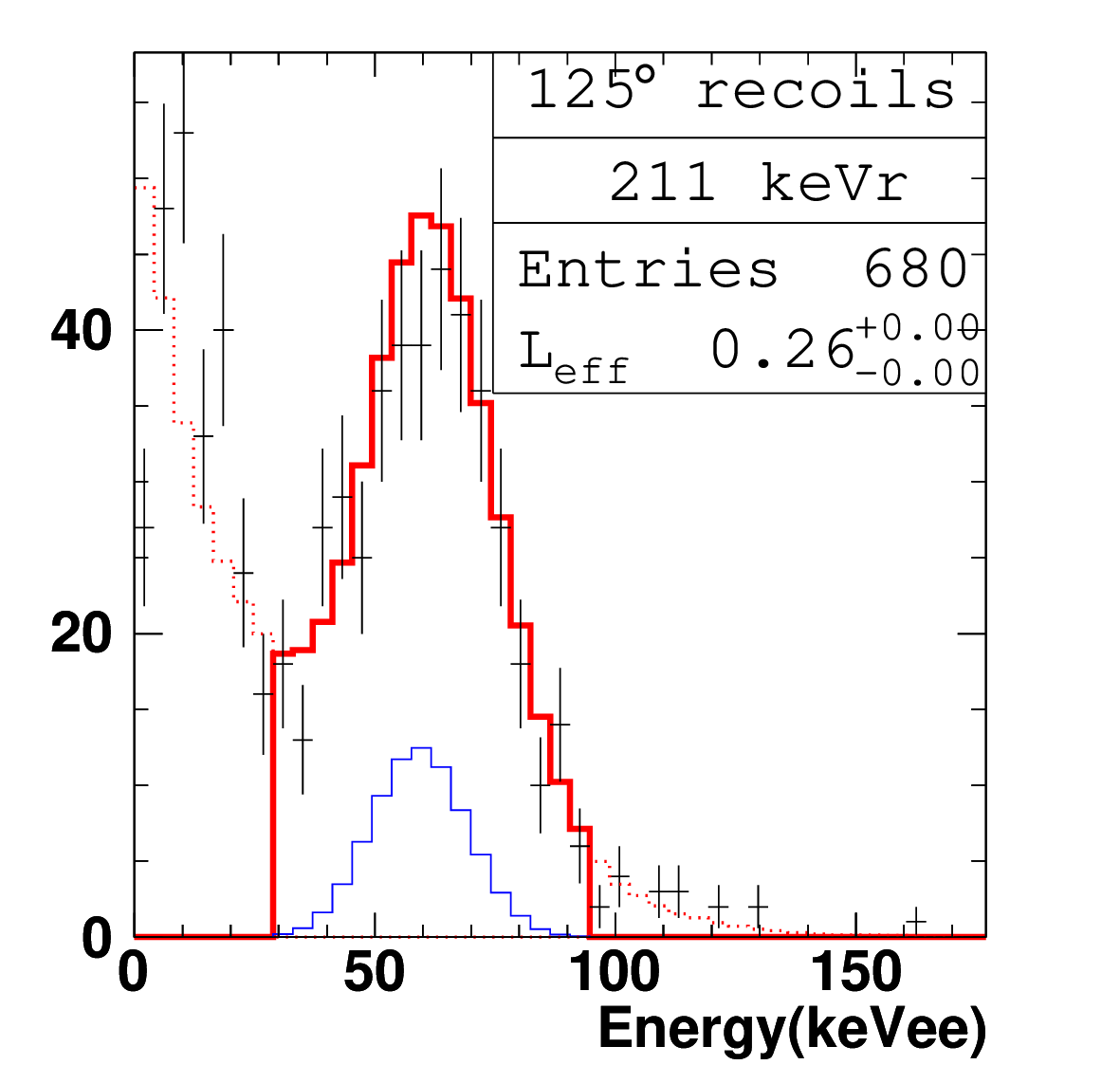}
    \includegraphics*[width=4.4cm]{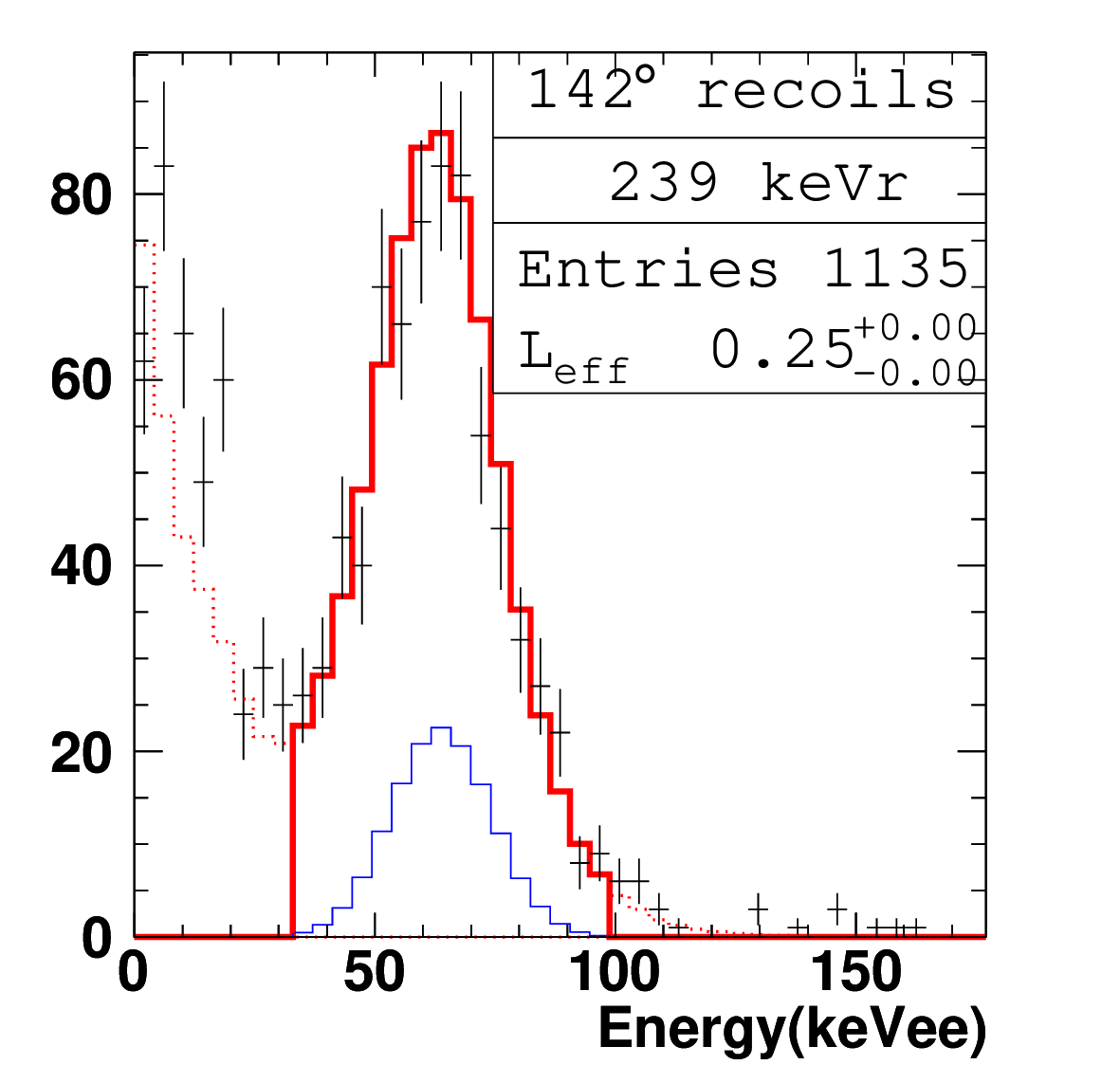}
    \caption{(Color online) 
Plotted are the recoil energy spectra for the 19 organic scintillator positions used in this experiment.
The data are taken with the organic scintillator located at the angle indicated in the legend of each plot with the
corresponding recoil energy indicated just below. The fit value for Leff with the statistical uncertainty from the fit is also listed in the legend of each plot. In each plot, the upper (red) histogram is the output of the GEANT4 based Monte Carlo
simulation of single and multiple neutron scatters in the detector. This upper histogram is fit to the data in the solid region, whereas the dotted part shows the MC simulation outside of the fit range. The lower solid histogram (blue) is the subset of the Monte Carlo events where the neutron only scatters once in the detector volume.}
  
    \label{fig:spectra}
  \end{center}
\end{figure*}

\begin{figure}[htbp]
  \begin{center}
    \includegraphics*[width=8cm]{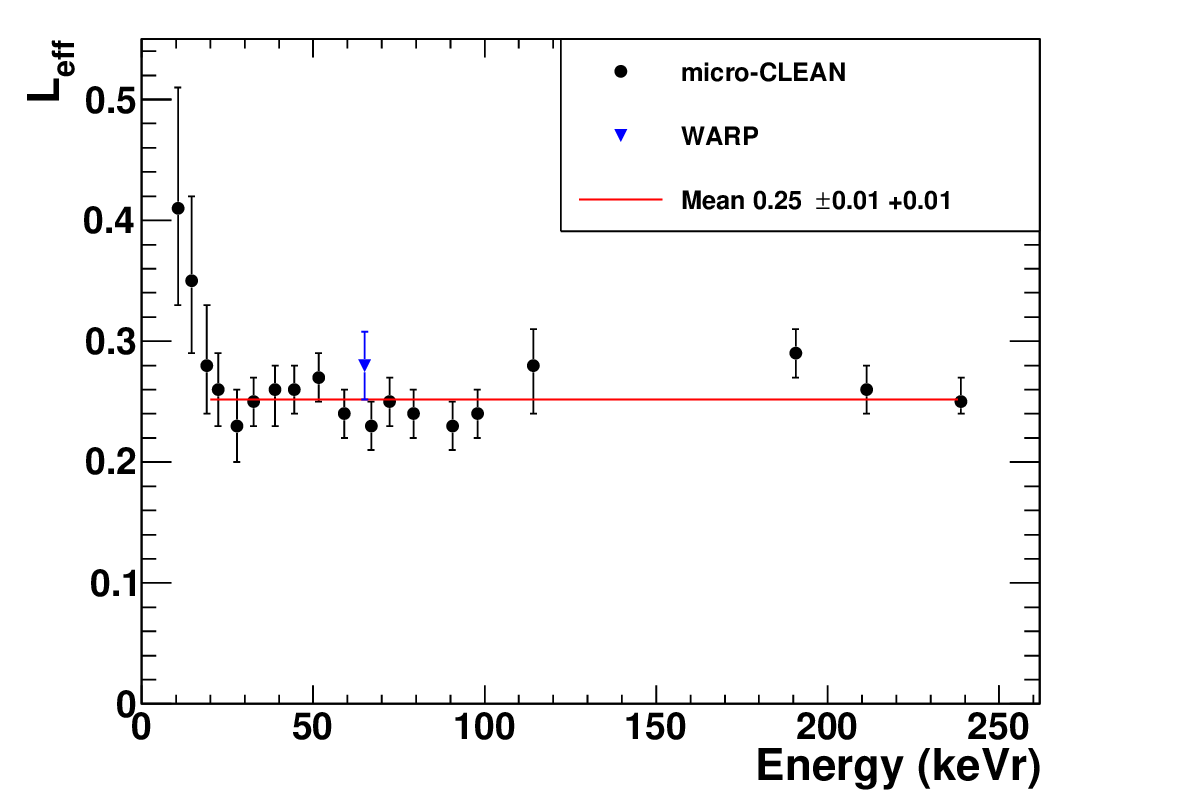}
    \caption{(Color online) Scintillation efficiency as a function of energy from 10 to 250
      keVr.  The weighted mean (red line) is generated from the data above 20 keVr and
      puts the mean scintillation efficiency at \Leff. The value measured by
      WARP is 0.28 at 65 keVr~\cite{Brunetti:2005}.} 
    \label{fig:quench}
  \end{center}
\end{figure}

The results for all organic scintillator positions are presented in
Fig.~\ref{fig:spectra}, and the scintillation efficiency as a function of
energy is shown in Fig.~\ref{fig:quench}. 
After studying the systematic effects described in the next section, we found 
that the individual scintillation efficiency values were constant across 
the range of recoil energies studied above 20 keVr, with a mean of \LeffErr.   
However, there still existed substantial differences between the simulation and the data, giving an average $\chi^{2}/d.o.f.$ of 3.7 for measurments above 20 keVr.
This disagreement is addressed in the next section.
Below 20 keVr, our data exhibit an upturn in scintillation efficiency as the 
energy goes to zero, and we were unable to find an experimental cause for 
this upturn. 
It is therefore unknown if this is a physically real effect or if we lose 
our ability to distinguish nuclear recoils from other backgrounds at these low
energies. 
All observed values and uncertainties are listed in Table~\ref{table:error_table}.

%%%%%%%%%%%%%%%%%%%%%%%%%%%%%%%%%%%%%%%%
\section{Systematic Uncertainties}
\label{sec:errors}
The sources of uncertainty considered are categorized as those associated with detector
operation, triggering effects, Monte Carlo background normalization, TOF window and fit range effects.
The uncertainties from the sources discussed in this section are all combined and the final resulting uncertainty for each scattering angle can be found in Table.~\ref{table:error_table}.

The first group of considered uncertainties deals with the data taking and stability of the neutron
and \Co~data runs.   Since the \Co~runs are used to calibrate the light yield
of the detector, the fit error of the \Co~peak and its stability over time directly
affect the measured scintillation efficiency. These  are estimated to be 2\% and 1.6\% respectively.
There is a second uncertainty in the angle of the organic
scintillator relative to the neutron generator that in turn introduces an
uncertainty in the corresponding energy via Eq.~\ref{eq:theta}. 
We have determined the uncertainty of the angular position of the organic 
scintillator to be $1.3^{\circ}$ at each position.

We examined the effects of the trigger efficiency, specifically looking
to address the upturn observed at low energies which could be explained by a
bias introduced by the trigger level. We took data for the
$22^{\circ}$ run at three different hardware triggers and the $26^{\circ}$
run at two different hardware triggers, and we
examined the effect of hardware and software triggers on both the
asymmetry cut and the final scintillation efficiency values. 
In all cases, the scintillation efficiency distributions did not
systematically change by varying the cuts and hardware threshold. We also
performed a toy Monte Carlo using the time dependence of the scintillation
light~\cite{Lippincott:2008} and the observed single photoelectron distribution to estimate possible
threshold effects. This study found the effect of any threshold bias given our
hardware trigger level to be less than 1\%, much smaller than the other errors in the
measurement. Therefore, we conclude that the triggering threshold does not explain the upturn at low energies.   

A third source of uncertainty arises because the Monte Carlo simulation, as mentioned at the end of Sec.~\ref{sec:analysis}, does not
exactly reproduce the observed background shape.  This can be seen by comparing the dashed and solid red lines in the large angle scattering spectra of Fig.~\ref{fig:spectra}.  
To account for this inconsistency, the data for each recoil energy were
reanalyzed under the assumption that the size of the multiple scattering background in the
histograms used to perform the fits varied by $\pm 50\%$
relative to that predicted by the Monte Carlo. The variations observed in this reanalysis 
are approximately $10\%$ below $20 $ keVr and $2\%$ above $20 $ keVr and
are included in the errors listed in Table~\ref{table:error_table}.

To determine the uncertainty due to the TOF cuts, the TOF window was expanded 
separately up and down in time by $50\%$.
This allows for recoil neutrons with smaller TOFs to be included when the window
is expanded downward and larger TOF neutrons when expanded upward.   
The effect of this variation was mostly in the lowest three data points and 
allowed them to move downwards in scintillation efficiency by about 0.04. 

There is an uncertainty associated with fitting the data in a limited range
around the predicted single-scattered neutron peak position. 
To estimate this uncertainty, we expand the fit range to include $\pm 5$ sigma
around the centroid of the single-scattered neutrons, instead of the 3-sigma
range used in the standard fit.   
The result of the wider fit range is to systematically push the determined scintillation efficiency up for energies between 20 keVr and 120 keVr.  
This effect appears to be caused by a disagreement in the high energy tails of the data and Monte Carlo, similar to the disagreement observed between Monte Carlo and data at lower energies for the high recoil angles. 
Changing the fit range adds a correlated error of $+0.01$ to the measured mean scintillation efficiency in Fig.~\ref{fig:quench}.

\begin{table}[ht]
  \centering
  \begin{tabular}{ccccc}\hline\hline
    Scattering angle ($^{\circ}$) & Energy (keVr) & \SE & + & - \\\hline
%    22       & 11	 & 0.41	 & 0.08	 & 0.08\\
    22       & 11	 & 0.41	 & 0.10	 & 0.08\\
%    26       & 15	 & 0.35	 & 0.05	 & 0.06\\
    26       & 15	 & 0.35	 & 0.07	 & 0.06\\
%    30       & 19	 & 0.28	 & 0.04	 & 0.04\\
    30       & 19	 & 0.28	 & 0.05	 & 0.04\\
    33       & 22	 & 0.26	 & 0.03	 & 0.03\\
    37       & 28	 & 0.23	 & 0.03	 & 0.03\\
    40       & 33	 & 0.25	 & 0.02	 & 0.02\\
    44       & 39	 & 0.26	 & 0.02	 & 0.03\\
    47       & 45	 & 0.26	 & 0.02	 & 0.02\\
    51       & 52	 & 0.27	 & 0.02	 & 0.02\\
    55       & 59	 & 0.24	 & 0.02	 & 0.02\\
    59       & 67	 & 0.23	 & 0.02	 & 0.02\\
    62       & 72	 & 0.25	 & 0.02	 & 0.02\\
    65       & 79	 & 0.24	 & 0.02	 & 0.02\\
    70       & 91	 & 0.23	 & 0.02	 & 0.02\\
    73       & 98	 & 0.24	 & 0.02	 & 0.02\\
    80       & 114	 & 0.28	 & 0.03	 & 0.04\\
    114       & 191	 & 0.29	 & 0.02	 & 0.02\\
    125       & 211	 & 0.26	 & 0.02	 & 0.02\\
    142       & 239	 & 0.25	 & 0.02	 & 0.01\\    \hline\hline
  \end{tabular}
  \caption{Table of energies and scintillation efficiencies from Fig. ~\ref{fig:quench}. 
    \SE\, values for energies above 20 keVr also have an additional correlated
    error of $+0.01$. 
The uncertainties shown are the combined statistical and systematic uncertainties.
} 
  \label{table:error_table}
\end{table}

\section{Results}
\label{sec:concl}
The scintillation efficiency for nuclear recoils in liquid argon has been
measured relative to electronic recoils for nuclear recoil energies from 10
keVr to 250 keVr. 
The scintillation efficiency values found at each of the recoil angles can be found in
Table~\ref{table:error_table} and are plotted in Fig.~\ref{fig:quench}.    
The ratio of the nuclear recoil scintillation response to the electronic recoil
response is \LeffErr~for recoils above 20 keVr.  
%The bin to bin correlation coefficient of the scintillation efficiency for the nuclear recoils above 20 keVr was 0.40.
An observed upturn in the scintillation efficiency below 20 keVr is currently unexplained. 
The scintillation efficiency for nuclear recoils should also have some dependence on any applied electric fields and this is a topic for further research.  

\section{Acknowledgements}
\label{sec:ack}
Computer resources were supplied by Yale University Biomedical High
Performance Computing Center and NIH grant: RR19895 and in part by the
facilities and staff of the Yale University Faculty of Arts and Sciences
High Performance Computing Center. 
We are also grateful for the support of the U.S. Department of Energy.
This work was supported by the David and Lucille Packard Foundation.


\begin{thebibliography}{29}
\expandafter\ifx\csname natexlab\endcsname\relax\def\natexlab#1{#1}\fi
\expandafter\ifx\csname bibnamefont\endcsname\relax
  \def\bibnamefont#1{#1}\fi
\expandafter\ifx\csname bibfnamefont\endcsname\relax
  \def\bibfnamefont#1{#1}\fi
\expandafter\ifx\csname citenamefont\endcsname\relax
  \def\citenamefont#1{#1}\fi
\expandafter\ifx\csname url\endcsname\relax
  \def\url#1{\texttt{#1}}\fi
\expandafter\ifx\csname urlprefix\endcsname\relax\def\urlprefix{URL }\fi
\providecommand{\bibinfo}[2]{#2}
\providecommand{\eprint}[2][]{\url{#2}}

\bibitem[{\citenamefont{Aprile et~al.}(2005{\natexlab{a}})\citenamefont{Aprile,
  Giboni, Majewski, Ni, Yamashita, Gaitskell, Sorensen, DeViveiros, Baudis,
  Bernstein et~al.}}]{Aprile:2005a}
\bibinfo{author}{\bibfnamefont{E.}~\bibnamefont{Aprile}},
  \bibinfo{author}{\bibfnamefont{K.~L.} \bibnamefont{Giboni}},
  \bibinfo{author}{\bibfnamefont{P.}~\bibnamefont{Majewski}},
  \bibinfo{author}{\bibfnamefont{K.}~\bibnamefont{Ni}},
  \bibinfo{author}{\bibfnamefont{M.}~\bibnamefont{Yamashita}},
  \bibinfo{author}{\bibfnamefont{R.}~\bibnamefont{Gaitskell}},
  \bibinfo{author}{\bibfnamefont{P.}~\bibnamefont{Sorensen}},
  \bibinfo{author}{\bibfnamefont{L.}~\bibnamefont{DeViveiros}},
  \bibinfo{author}{\bibfnamefont{L.}~\bibnamefont{Baudis}},
  \bibinfo{author}{\bibfnamefont{A.}~\bibnamefont{Bernstein}},
  \bibnamefont{et~al.}, \bibinfo{journal}{New Ast. Rev.}
  \textbf{\bibinfo{volume}{49}}, \bibinfo{pages}{289}
  (\bibinfo{year}{2005}{\natexlab{a}}).

\bibitem[{\citenamefont{Cline et~al.}(2003)\citenamefont{Cline, Seo,
  Sergiampietri, Wang, White, Gao, Picchi, Mannocchi, Periale, Pietropaolo
  et~al.}}]{Cline:2003}
\bibinfo{author}{\bibfnamefont{D.~B.} \bibnamefont{Cline}},
  \bibinfo{author}{\bibfnamefont{Y.}~\bibnamefont{Seo}},
  \bibinfo{author}{\bibfnamefont{F.}~\bibnamefont{Sergiampietri}},
  \bibinfo{author}{\bibfnamefont{H.}~\bibnamefont{Wang}},
  \bibinfo{author}{\bibfnamefont{J.~T.} \bibnamefont{White}},
  \bibinfo{author}{\bibfnamefont{J.}~\bibnamefont{Gao}},
  \bibinfo{author}{\bibfnamefont{P.}~\bibnamefont{Picchi}},
  \bibinfo{author}{\bibfnamefont{G.}~\bibnamefont{Mannocchi}},
  \bibinfo{author}{\bibfnamefont{L.}~\bibnamefont{Periale}},
  \bibinfo{author}{\bibfnamefont{F.}~\bibnamefont{Pietropaolo}},
  \bibnamefont{et~al.}, \bibinfo{journal}{Nucl. Phys. B (Proc. Suppl.)}
  \textbf{\bibinfo{volume}{124}}, \bibinfo{pages}{229} (\bibinfo{year}{2003}).

\bibitem[{\citenamefont{Brunetti et~al.}(2005)\citenamefont{Brunetti,
  Calligarich, Cambiaghi, Carbonara, Cocco, De~Vecchi, Dolfini, Ereditato,
  Fiorillo, Grandi et~al.}}]{Brunetti:2005}
\bibinfo{author}{\bibfnamefont{R.}~\bibnamefont{Brunetti}},
  \bibinfo{author}{\bibfnamefont{E.}~\bibnamefont{Calligarich}},
  \bibinfo{author}{\bibfnamefont{M.}~\bibnamefont{Cambiaghi}},
  \bibinfo{author}{\bibfnamefont{F.}~\bibnamefont{Carbonara}},
  \bibinfo{author}{\bibfnamefont{A.}~\bibnamefont{Cocco}},
  \bibinfo{author}{\bibfnamefont{C.}~\bibnamefont{De~Vecchi}},
  \bibinfo{author}{\bibfnamefont{R.}~\bibnamefont{Dolfini}},
  \bibinfo{author}{\bibfnamefont{A.}~\bibnamefont{Ereditato}},
  \bibinfo{author}{\bibfnamefont{G.}~\bibnamefont{Fiorillo}},
  \bibinfo{author}{\bibfnamefont{L.}~\bibnamefont{Grandi}},
  \bibnamefont{et~al.}, \bibinfo{journal}{New Ast. Rev.}
  \textbf{\bibinfo{volume}{49}}, \bibinfo{pages}{265} (\bibinfo{year}{2005}).

\bibitem[{\citenamefont{Benetti et~al.}(2007)\citenamefont{Benetti, Acciarri,
  Adamo, Baibussinov, Baldo-Ceolin, Belluco, Calaprice, Calligarich, Cambiaghi,
  Carbonara et~al.}}]{Benetti:2007}
\bibinfo{author}{\bibfnamefont{P.}~\bibnamefont{Benetti}},
  \bibinfo{author}{\bibfnamefont{R.}~\bibnamefont{Acciarri}},
  \bibinfo{author}{\bibfnamefont{F.}~\bibnamefont{Adamo}},
  \bibinfo{author}{\bibfnamefont{B.}~\bibnamefont{Baibussinov}},
  \bibinfo{author}{\bibfnamefont{M.}~\bibnamefont{Baldo-Ceolin}},
  \bibinfo{author}{\bibfnamefont{M.}~\bibnamefont{Belluco}},
  \bibinfo{author}{\bibfnamefont{F.}~\bibnamefont{Calaprice}},
  \bibinfo{author}{\bibfnamefont{E.}~\bibnamefont{Calligarich}},
  \bibinfo{author}{\bibfnamefont{M.}~\bibnamefont{Cambiaghi}},
  \bibinfo{author}{\bibfnamefont{F.}~\bibnamefont{Carbonara}},
  \bibnamefont{et~al.}, \bibinfo{journal}{arXiv:astro-ph/0701286}
  (\bibinfo{year}{2007}).

\bibitem[{\citenamefont{Rubbia}(2006)}]{Rubbia:2006}
\bibinfo{author}{\bibfnamefont{A.}~\bibnamefont{Rubbia}}, \bibinfo{journal}{J.
  Phys.: Conf. Ser.} \textbf{\bibinfo{volume}{39}}, \bibinfo{pages}{129}
  (\bibinfo{year}{2006}).

\bibitem[{\citenamefont{Jungman et~al.}(1996)\citenamefont{Jungman,
  Kamionkowski, and Griest}}]{Jungman:1996}
\bibinfo{author}{\bibfnamefont{G.}~\bibnamefont{Jungman}},
  \bibinfo{author}{\bibfnamefont{M.}~\bibnamefont{Kamionkowski}},
  \bibnamefont{and} \bibinfo{author}{\bibfnamefont{K.}~\bibnamefont{Griest}},
  \bibinfo{journal}{Phys. Rept.} \textbf{\bibinfo{volume}{267}},
  \bibinfo{pages}{195} (\bibinfo{year}{1996}).

\bibitem[{\citenamefont{Ahmed et~al.}(2009)\citenamefont{Ahmed, Akerib,
  Arrenberg, Bailey, Balakishiyeva, Baudis, Bauer, Brink, Bruch, Bunker
  et~al.}}]{Ahmed:2009}
\bibinfo{author}{\bibfnamefont{Z.}~\bibnamefont{Ahmed}},
  \bibinfo{author}{\bibfnamefont{D.}~\bibnamefont{Akerib}},
  \bibinfo{author}{\bibfnamefont{S.}~\bibnamefont{Arrenberg}},
  \bibinfo{author}{\bibfnamefont{C.}~\bibnamefont{Bailey}},
  \bibinfo{author}{\bibfnamefont{D.}~\bibnamefont{Balakishiyeva}},
  \bibinfo{author}{\bibfnamefont{L.}~\bibnamefont{Baudis}},
  \bibinfo{author}{\bibfnamefont{D.}~\bibnamefont{Bauer}},
  \bibinfo{author}{\bibfnamefont{P.}~\bibnamefont{Brink}},
  \bibinfo{author}{\bibfnamefont{T.}~\bibnamefont{Bruch}},
  \bibinfo{author}{\bibfnamefont{R.}~\bibnamefont{Bunker}},
  \bibnamefont{et~al.} (\bibinfo{collaboration}{CDMS}) (\bibinfo{year}{2009}),
  \eprint{0912.3592}.

\bibitem[{\citenamefont{Angle et~al.}(2007)\citenamefont{Angle, Aprile,
  Arneodo, Baudis, Bernstein, Bolozdynya, Brusov, Coelho, Dahl, DeViveiros
  et~al.}}]{Angle:2007}
\bibinfo{author}{\bibfnamefont{J.}~\bibnamefont{Angle}},
  \bibinfo{author}{\bibfnamefont{E.}~\bibnamefont{Aprile}},
  \bibinfo{author}{\bibfnamefont{F.}~\bibnamefont{Arneodo}},
  \bibinfo{author}{\bibfnamefont{L.}~\bibnamefont{Baudis}},
  \bibinfo{author}{\bibfnamefont{A.}~\bibnamefont{Bernstein}},
  \bibinfo{author}{\bibfnamefont{A.}~\bibnamefont{Bolozdynya}},
  \bibinfo{author}{\bibfnamefont{P.}~\bibnamefont{Brusov}},
  \bibinfo{author}{\bibfnamefont{L.}~\bibnamefont{Coelho}},
  \bibinfo{author}{\bibfnamefont{C.}~\bibnamefont{Dahl}},
  \bibinfo{author}{\bibfnamefont{L.}~\bibnamefont{DeViveiros}},
  \bibnamefont{et~al.}, \bibinfo{journal}{arXiv:0706.0039}
  (\bibinfo{year}{2007}).

\bibitem[{\citenamefont{Hitachi et~al.}(1983)\citenamefont{Hitachi, Takahashi,
  Funayama, Masuda, Kikuchi, and Doke}}]{Hitachi:1983}
\bibinfo{author}{\bibfnamefont{A.}~\bibnamefont{Hitachi}},
  \bibinfo{author}{\bibfnamefont{T.}~\bibnamefont{Takahashi}},
  \bibinfo{author}{\bibfnamefont{N.}~\bibnamefont{Funayama}},
  \bibinfo{author}{\bibfnamefont{K.}~\bibnamefont{Masuda}},
  \bibinfo{author}{\bibfnamefont{J.}~\bibnamefont{Kikuchi}}, \bibnamefont{and}
  \bibinfo{author}{\bibfnamefont{T.}~\bibnamefont{Doke}},
  \bibinfo{journal}{Physical Review B} \textbf{\bibinfo{volume}{27}},
  \bibinfo{pages}{5279} (\bibinfo{year}{1983}).

\bibitem[{\citenamefont{Lippincott et~al.}(2008)\citenamefont{Lippincott,
  McKinsey, Nikkel, Coakley, Hime, Stonehill, Gastler, and
  Kearns}}]{Lippincott:2008}
\bibinfo{author}{\bibfnamefont{W.}~\bibnamefont{Lippincott}},
  \bibinfo{author}{\bibfnamefont{D.}~\bibnamefont{McKinsey}},
  \bibinfo{author}{\bibfnamefont{J.}~\bibnamefont{Nikkel}},
  \bibinfo{author}{\bibfnamefont{K.}~\bibnamefont{Coakley}},
  \bibinfo{author}{\bibfnamefont{A.}~\bibnamefont{Hime}},
  \bibinfo{author}{\bibfnamefont{L.}~\bibnamefont{Stonehill}},
  \bibinfo{author}{\bibfnamefont{D.}~\bibnamefont{Gastler}}, \bibnamefont{and}
  \bibinfo{author}{\bibfnamefont{E.}~\bibnamefont{Kearns}},
  \bibinfo{journal}{Phys.Rev.C78:035801}  (\bibinfo{year}{2008}).

\bibitem[{\citenamefont{Takahashi et~al.}(1975)\citenamefont{Takahashi, Konno,
  Hamada, Miyajima, Kubota, Nakamoto, Hitachi, Shibamura, and
  Doke}}]{Takahashi:1975}
\bibinfo{author}{\bibfnamefont{T.}~\bibnamefont{Takahashi}},
  \bibinfo{author}{\bibfnamefont{S.}~\bibnamefont{Konno}},
  \bibinfo{author}{\bibfnamefont{T.}~\bibnamefont{Hamada}},
  \bibinfo{author}{\bibfnamefont{M.}~\bibnamefont{Miyajima}},
  \bibinfo{author}{\bibfnamefont{S.}~\bibnamefont{Kubota}},
  \bibinfo{author}{\bibfnamefont{S.}~\bibnamefont{Nakamoto}},
  \bibinfo{author}{\bibfnamefont{A.}~\bibnamefont{Hitachi}},
  \bibinfo{author}{\bibfnamefont{E.}~\bibnamefont{Shibamura}},
  \bibnamefont{and} \bibinfo{author}{\bibfnamefont{T.}~\bibnamefont{Doke}},
  \bibinfo{journal}{Phys. Rev. A} \textbf{\bibinfo{volume}{12}},
  \bibinfo{pages}{1771} (\bibinfo{year}{1975}).

\bibitem[{\citenamefont{Miyajima et~al.}(1974)\citenamefont{Miyajima,
  Takahashi, Konno, Hamada, Kubota, Shibamura, and Doke}}]{Miyajima:1974}
\bibinfo{author}{\bibfnamefont{M.}~\bibnamefont{Miyajima}},
  \bibinfo{author}{\bibfnamefont{T.}~\bibnamefont{Takahashi}},
  \bibinfo{author}{\bibfnamefont{S.}~\bibnamefont{Konno}},
  \bibinfo{author}{\bibfnamefont{T.}~\bibnamefont{Hamada}},
  \bibinfo{author}{\bibfnamefont{S.}~\bibnamefont{Kubota}},
  \bibinfo{author}{\bibfnamefont{H.}~\bibnamefont{Shibamura}},
  \bibnamefont{and} \bibinfo{author}{\bibfnamefont{T.}~\bibnamefont{Doke}},
  \bibinfo{journal}{Phys. Rev. A} \textbf{\bibinfo{volume}{9}},
  \bibinfo{pages}{1438} (\bibinfo{year}{1974}).

\bibitem[{\citenamefont{Doke et~al.}(2002)\citenamefont{Doke, Hitachi, Kikuchi,
  Masuda, Okada, and Shibamura}}]{Doke:2002}
\bibinfo{author}{\bibfnamefont{T.}~\bibnamefont{Doke}},
  \bibinfo{author}{\bibfnamefont{A.}~\bibnamefont{Hitachi}},
  \bibinfo{author}{\bibfnamefont{J.}~\bibnamefont{Kikuchi}},
  \bibinfo{author}{\bibfnamefont{K.}~\bibnamefont{Masuda}},
  \bibinfo{author}{\bibfnamefont{H.}~\bibnamefont{Okada}}, \bibnamefont{and}
  \bibinfo{author}{\bibfnamefont{E.}~\bibnamefont{Shibamura}},
  \bibinfo{journal}{Jpn. J. Appl. Phys.} \textbf{\bibinfo{volume}{41}},
  \bibinfo{pages}{1538} (\bibinfo{year}{2002}).

\bibitem[{\citenamefont{Lindhard et~al.}(1963)\citenamefont{Lindhard, Scharff,
  and Schi{\o}tt}}]{Lindhard:1963}
\bibinfo{author}{\bibfnamefont{J.}~\bibnamefont{Lindhard}},
  \bibinfo{author}{\bibfnamefont{M.}~\bibnamefont{Scharff}}, \bibnamefont{and}
  \bibinfo{author}{\bibfnamefont{H.}~\bibnamefont{Schi{\o}tt}},
  \bibinfo{journal}{Mat. Fys. Medd. K. Dan. Vidensk. Selsk.}
  \textbf{\bibinfo{volume}{33}}, \bibinfo{pages}{1} (\bibinfo{year}{1963}).

\bibitem[{\citenamefont{Sorensen et~al.}(2009)}]{Sorensen:2008ec}
\bibinfo{author}{\bibfnamefont{P.}~\bibnamefont{Sorensen}}
  \bibnamefont{et~al.}, \bibinfo{journal}{Nucl. Instrum. Meth.}
  \textbf{\bibinfo{volume}{A601}}, \bibinfo{pages}{339} (\bibinfo{year}{2009}),
  \eprint{0807.0459}.

\bibitem[{\citenamefont{Aprile et~al.}(2009)}]{Aprile:2008rc}
\bibinfo{author}{\bibfnamefont{E.}~\bibnamefont{Aprile}} \bibnamefont{et~al.},
  \bibinfo{journal}{Phys. Rev.} \textbf{\bibinfo{volume}{C79}},
  \bibinfo{pages}{045807} (\bibinfo{year}{2009}), \eprint{0810.0274}.

\bibitem[{\citenamefont{Aprile et~al.}(2005{\natexlab{b}})\citenamefont{Aprile,
  Giboni, Majewski, Ni, Yamashita, Hasty, Manzur, and McKinsey}}]{Aprile:2005}
\bibinfo{author}{\bibfnamefont{E.}~\bibnamefont{Aprile}},
  \bibinfo{author}{\bibfnamefont{K.~L.} \bibnamefont{Giboni}},
  \bibinfo{author}{\bibfnamefont{P.}~\bibnamefont{Majewski}},
  \bibinfo{author}{\bibfnamefont{K.}~\bibnamefont{Ni}},
  \bibinfo{author}{\bibfnamefont{M.}~\bibnamefont{Yamashita}},
  \bibinfo{author}{\bibfnamefont{R.}~\bibnamefont{Hasty}},
  \bibinfo{author}{\bibfnamefont{A.}~\bibnamefont{Manzur}}, \bibnamefont{and}
  \bibinfo{author}{\bibfnamefont{D.~N.} \bibnamefont{McKinsey}},
  \bibinfo{journal}{Phys. Rev. D} \textbf{\bibinfo{volume}{72}},
  \bibinfo{pages}{72006} (\bibinfo{year}{2005}{\natexlab{b}}).

\bibitem[{\citenamefont{Chepel et~al.}(2006)\citenamefont{Chepel, Solovov,
  Neves, Pereira, Mendes, Silva, Lindote, Pinto~da Cunha, Lopes, and
  Kossionides}}]{Chepel:2006}
\bibinfo{author}{\bibfnamefont{V.}~\bibnamefont{Chepel}},
  \bibinfo{author}{\bibfnamefont{V.}~\bibnamefont{Solovov}},
  \bibinfo{author}{\bibfnamefont{F.}~\bibnamefont{Neves}},
  \bibinfo{author}{\bibfnamefont{A.}~\bibnamefont{Pereira}},
  \bibinfo{author}{\bibfnamefont{P.~J.} \bibnamefont{Mendes}},
  \bibinfo{author}{\bibfnamefont{C.~P.} \bibnamefont{Silva}},
  \bibinfo{author}{\bibfnamefont{A.}~\bibnamefont{Lindote}},
  \bibinfo{author}{\bibfnamefont{J.}~\bibnamefont{Pinto~da Cunha}},
  \bibinfo{author}{\bibfnamefont{M.~I.} \bibnamefont{Lopes}}, \bibnamefont{and}
  \bibinfo{author}{\bibfnamefont{S.}~\bibnamefont{Kossionides}},
  \bibinfo{journal}{Astropart. Phys.} \textbf{\bibinfo{volume}{26}},
  \bibinfo{pages}{58} (\bibinfo{year}{2006}).

\bibitem[{\citenamefont{Manzur et~al.}()\citenamefont{Manzur, Curioni, Kastens,
  McKinsey, Ni, and Wongjirad}}]{Manzur:2009}
\bibinfo{author}{\bibfnamefont{A.}~\bibnamefont{Manzur}},
  \bibinfo{author}{\bibfnamefont{A.}~\bibnamefont{Curioni}},
  \bibinfo{author}{\bibfnamefont{L.}~\bibnamefont{Kastens}},
  \bibinfo{author}{\bibfnamefont{D.}~\bibnamefont{McKinsey}},
  \bibinfo{author}{\bibfnamefont{K.}~\bibnamefont{Ni}}, \bibnamefont{and}
  \bibinfo{author}{\bibfnamefont{T.}~\bibnamefont{Wongjirad}},
  \urlprefix\url{http://arxiv.org/abs/0909.1063v4}.

\bibitem[{\citenamefont{Hitachi et~al.}(1992)\citenamefont{Hitachi, Doke, and
  Mozumder}}]{Hitachi:1992}
\bibinfo{author}{\bibfnamefont{A.}~\bibnamefont{Hitachi}},
  \bibinfo{author}{\bibfnamefont{T.}~\bibnamefont{Doke}}, \bibnamefont{and}
  \bibinfo{author}{\bibfnamefont{A.}~\bibnamefont{Mozumder}},
  \bibinfo{journal}{Phys. Rev. B} \textbf{\bibinfo{volume}{46}},
  \bibinfo{pages}{11463} (\bibinfo{year}{1992}).

\bibitem[{\citenamefont{Mei et~al.}(2008)\citenamefont{Mei, Yin, Stonehill, and
  Hime}}]{Mei:2008}
\bibinfo{author}{\bibfnamefont{D.-M.} \bibnamefont{Mei}},
  \bibinfo{author}{\bibfnamefont{Z.}~\bibnamefont{Yin}},
  \bibinfo{author}{\bibfnamefont{L.}~\bibnamefont{Stonehill}},
  \bibnamefont{and} \bibinfo{author}{\bibfnamefont{A.}~\bibnamefont{Hime}},
  \bibinfo{journal}{Astropart.Phys.30:12-17}  (\bibinfo{year}{2008}).

\bibitem[{\citenamefont{Hitachi et~al.}(1982)\citenamefont{Hitachi, Takahashi,
  Hamada, Shibamura, Funayama, Masuda, Kikuchi, and Doke}}]{Hitachi:1982}
\bibinfo{author}{\bibfnamefont{A.}~\bibnamefont{Hitachi}},
  \bibinfo{author}{\bibfnamefont{T.}~\bibnamefont{Takahashi}},
  \bibinfo{author}{\bibfnamefont{T.}~\bibnamefont{Hamada}},
  \bibinfo{author}{\bibfnamefont{E.}~\bibnamefont{Shibamura}},
  \bibinfo{author}{\bibfnamefont{N.}~\bibnamefont{Funayama}},
  \bibinfo{author}{\bibfnamefont{K.}~\bibnamefont{Masuda}},
  \bibinfo{author}{\bibfnamefont{J.}~\bibnamefont{Kikuchi}}, \bibnamefont{and}
  \bibinfo{author}{\bibfnamefont{T.}~\bibnamefont{Doke}},
  \bibinfo{journal}{Nucl. Instrum. Meth.} \textbf{\bibinfo{volume}{196}},
  \bibinfo{pages}{97} (\bibinfo{year}{1982}).

\bibitem[{\citenamefont{Regenfus}(2007)}]{Regenfus:2007}
\bibinfo{author}{\bibfnamefont{C.}~\bibnamefont{Regenfus}},
  \bibinfo{journal}{IDM2006 Proceedings, World Scientific}
  \textbf{\bibinfo{volume}{15}}, \bibinfo{pages}{32} (\bibinfo{year}{2007}).

\bibitem[{\citenamefont{Cheshnovsky et~al.}(1972)\citenamefont{Cheshnovsky,
  Raz, and Jortner}}]{Cheshnovsky:1972}
\bibinfo{author}{\bibfnamefont{O.}~\bibnamefont{Cheshnovsky}},
  \bibinfo{author}{\bibfnamefont{B.}~\bibnamefont{Raz}}, \bibnamefont{and}
  \bibinfo{author}{\bibfnamefont{J.}~\bibnamefont{Jortner}},
  \bibinfo{journal}{J. Chem. Phys.} \textbf{\bibinfo{volume}{57}},
  \bibinfo{pages}{4628} (\bibinfo{year}{1972}).

\bibitem[{\citenamefont{McKinsey et~al.}(1997)\citenamefont{McKinsey, Brome,
  Butterworth, Golub, Habicht, Huffman, Lamoreaux, Mattoni, and
  Doyle}}]{McKinsey:1997}
\bibinfo{author}{\bibfnamefont{D.~N.} \bibnamefont{McKinsey}},
  \bibinfo{author}{\bibfnamefont{C.~R.} \bibnamefont{Brome}},
  \bibinfo{author}{\bibfnamefont{J.~S.} \bibnamefont{Butterworth}},
  \bibinfo{author}{\bibfnamefont{R.}~\bibnamefont{Golub}},
  \bibinfo{author}{\bibfnamefont{K.}~\bibnamefont{Habicht}},
  \bibinfo{author}{\bibfnamefont{P.~R.} \bibnamefont{Huffman}},
  \bibinfo{author}{\bibfnamefont{S.~K.} \bibnamefont{Lamoreaux}},
  \bibinfo{author}{\bibfnamefont{C.~E.~H.} \bibnamefont{Mattoni}},
  \bibnamefont{and} \bibinfo{author}{\bibfnamefont{J.~M.} \bibnamefont{Doyle}},
  \bibinfo{journal}{Nucl. Instrum. Meth. B} \textbf{\bibinfo{volume}{132}},
  \bibinfo{pages}{351} (\bibinfo{year}{1997}).

\bibitem[{GEA()}]{GEANT}
\urlprefix\url{http://geant4.cern.ch/}.

\bibitem[{CLH()}]{CLHEP}
\urlprefix\url{http://proj-clhep.web.cern.ch/}.

\bibitem[{ROO()}]{ROOT}
\urlprefix\url{http://root.cern.ch/}.

\bibitem[{\citenamefont{Conti et~al.}(2003)\citenamefont{Conti, DeVoe, Gratta,
  Koffas, Waldman, Wodin, Akimov, Bower, Breidenbach, Conley
  et~al.}}]{Conti:2003}
\bibinfo{author}{\bibfnamefont{E.}~\bibnamefont{Conti}},
  \bibinfo{author}{\bibfnamefont{R.}~\bibnamefont{DeVoe}},
  \bibinfo{author}{\bibfnamefont{G.}~\bibnamefont{Gratta}},
  \bibinfo{author}{\bibfnamefont{T.}~\bibnamefont{Koffas}},
  \bibinfo{author}{\bibfnamefont{S.}~\bibnamefont{Waldman}},
  \bibinfo{author}{\bibfnamefont{J.}~\bibnamefont{Wodin}},
  \bibinfo{author}{\bibfnamefont{D.}~\bibnamefont{Akimov}},
  \bibinfo{author}{\bibfnamefont{G.}~\bibnamefont{Bower}},
  \bibinfo{author}{\bibfnamefont{M.}~\bibnamefont{Breidenbach}},
  \bibinfo{author}{\bibfnamefont{R.}~\bibnamefont{Conley}},
  \bibnamefont{et~al.}, \bibinfo{journal}{Phys.Rev.B68:054201,2003}
  (\bibinfo{year}{2003}).

\end{thebibliography}
\end{document}